%% file: ms.tex
\documentclass[10pt, format=manuscript,nonacm]{acmart}
\usepackage[utf8]{inputenc}

\input{utils/my_commands.tex}

\title{CC-Light eQASM Architecture Specification}
\author{X.~Fu}
\email{xiangfu@quanta.org.cn}
\affiliation{%
  \institution{QuTech, Delft University of Technology}
}

\date{May 2018}

\begin{document}

\begin{abstract}
This document~\footnote{The CC-Light eQASM is a joint effort of the team (X.~Fu, L.~Riesebos, J.~van~Someren, and J.~van~Straten) in QuTech. This specification is written by X.~Fu. If eQASM or CC-Light is used in your work, please cite to the following paper: \\ \indent \tiny{X.~Fu, L.~Riesebos, M.~A.~Rol, J.~van Straten, J.~van~Someren, N.~Khammassi, I.~Ashraf, R.~F.~L.~Vermeulen, V.~Newsum, K.~K.~L.~Loh, J.~C.~de~Sterke, W.~J.~Vlothuizen, R.~N.~Schouten, C.~G.~Almudever, L.~DiCarlo, and K.~Bertels, \textbf{eQASM: An Executable Quantum Instruction Set Architecture}, \textit{IEEE International Symposium on High Performance ComputerArchitecture (HPCA)}, pp.224-237, IEEE, 2019.}} is the specification of the CC-Light instantiation of  executable QASM (eQASM), a quantum instruction set architecture (QISA) developed in QuTech targeting to control a seven-qubit superconducting quantum processor. This document can serve as a reference manual for low-level programmers, compiler backend developers, and microarchitecture implementers of eQASM. The design of CC-Light eQASM is under the \href{https://www.apache.org/licenses/LICENSE-2.0}{Apache 2.0 License}.

\end{abstract}
\maketitle
\newpage
\tableofcontents

\input{00-intro.tex}
\input{01-overview.tex}
\input{02-eqasm_syntax.tex}
\input{03-isa.tex}

\bibliographystyle{IEEEtran}
\bibliography{reference}

\newpage
\appendix
\input{app-01-examples.tex}

\input{app-02-qmap.tex}

\end{document}

%% file: utils/my_commands.tex
\usepackage{fancyhdr}
\usepackage{hyperref}

\usepackage{amssymb,amsmath}
\usepackage{graphicx}

\usepackage[ruled]{algorithm2e}

\usepackage{xcolor}

\usepackage{listings}

\usepackage{vhistory}

\usepackage{verbatim}

\usepackage{multirow}

\usepackage{braket}

\usepackage{array}

\usepackage{tikz}
\usetikzlibrary{calc}
\usetikzlibrary{backgrounds}
\usetikzlibrary{math}
\usetikzlibrary{arrows.meta}

\usepackage{tablefootnote}

\usepackage{enumitem}

\usepackage{changepage}

\input{utils/Qcircuit.tex}

\definecolor{eclipseBlue}{RGB}{42,0.0,255}
\definecolor{eclipseGreen}{RGB}{63,127,95}
\definecolor{eclipsePurple}{RGB}{127,0,85}
\definecolor{eclipseRed}{RGB}{223,26,65}
\definecolor{sublimegreen}{RGB}{112,200,0}
\definecolor{purple}{RGB}{255, 0, 255}

\definecolor{grayonine}{gray}{0.9}
\definecolor{grayoeight}{gray}{0.8}
\definecolor{grayoseven}{gray}{0.7}
\definecolor{grayosix}{gray}{0.6}
\definecolor{grayofive}{gray}{0.5}
\definecolor{grayofour}{gray}{0.4}
\definecolor{grayothree}{gray}{0.3}
\definecolor{grayotwo}{gray}{0.2}
\definecolor{grayoone}{gray}{0.1}

\definecolor{opcodecolor}{RGB}{0,0,0}

\newcommand{\filename}[1]{\colorbox{grayonine}{\small \texttt{#1}}}
\newcommand{\code}[1]{{\small \texttt{#1}}}

\newcommand{\us}[1]{$#1~\mu\mathrm{s}$}

\newcommand{\bits}[1]{$#1~\mathrm{bits}$}

\newcommand{\bin}[1]{{`\texttt{#1}'}}

\newcommand{\qgate}[1]{$#1$}

\DeclareFontFamily{U}{matha}{\hyphenchar\font45}
\DeclareFontShape{U}{matha}{m}{n}{
      <5> <6> <7> <8> <9> <10> gen * matha
      <10.95> matha10 <12> <14.4> <17.28> <20.74> <24.88> matha12
      }{}
\DeclareSymbolFont{matha}{U}{matha}{m}{n}
\DeclareMathSymbol{\wedge}         {2}{matha}{"5E}
\DeclareMathSymbol{\vee}           {2}{matha}{"5F}

\usepackage{mathtools}

\DeclarePairedDelimiter\floor{\lfloor}{\rfloor}

\makeatletter
\lst@Key{matchrangestart}{f}{\lstKV@SetIf{#1}\lst@ifmatchrangestart}
\def\lst@SkipToFirst{%
    \lst@ifmatchrangestart\c@lstnumber=\numexpr-1+\lst@firstline\fi
    \ifnum \lst@lineno<\lst@firstline
        \def\lst@next{\lst@BeginDropInput\lst@Pmode
        \lst@Let{13}\lst@MSkipToFirst
        \lst@Let{10}\lst@MSkipToFirst}%
        \expandafter\lst@next
    \else
        \expandafter\lst@BOLGobble
    \fi}
\makeatother

\input{utils/eQASM_lang.tex}
\input{utils/behaviordescription.tex}
\input{utils/insn_table_header.tex}

\newcolumntype{L}[1]{>{\raggedright\let\newline\\\arraybackslash\hspace{0pt}}m{#1}}
\newcolumntype{C}[1]{>{\centering\let\newline\\\arraybackslash\hspace{0pt}}m{#1}}
\newcolumntype{R}[1]{>{\raggedleft\let\newline\\\arraybackslash\hspace{0pt}}m{#1}}

\usepackage{tikz,colortbl}
\usetikzlibrary{calc}
\usepackage{zref-savepos}

\newcounter{NoTableEntry}
\renewcommand*{\theNoTableEntry}{NTE-\the\value{NoTableEntry}}

\providecommand{\tightlist}{\setlength{\itemsep}{0pt}\setlength{\parskip}{0pt}}


%% file: utils/eQASM_lang.tex
\usepackage{listings}

\lstdefinelanguage{eQASM}
{
  sensitive=false, 
  alsoletter={.1234567890},
  keywords={},
  otherkeywords={ 
  |
  },
  morekeywords=[2]{always, never, eq, ne, eqz, nez, lt, ltz, le, gt, ge, gez, ltu, leu, gtu, geu, carry, notcarry},
  morekeywords=[3]{qwait, qwaitr, SMIS, SMIT},
  morekeywords=[4]{bs},
  morekeywords=[5]{MeasZ, X, Y, CNOT, X180, Xm180, X90, Xm90, Y180, Ym180, Ym90, Y90, CZ, T, H, S, Z, QNOP},
  morekeywords=[6]{nop, stop, ldi, ldui, add, sub, addc, subc, and, or, xor, not, fmr, br, cmp, fbr,  st, ld},
  morekeywords=[7]{.register, .def_sym},
  morekeywords=[8]{nand, nor, xnor, bra, brn, beq, bne, blt, ble, bgt, bge, bltu, bleu, bgtu, bgeu, goto, mov, shl1, mult2},
  morecomment=[l]{\#}, 
}

\lstdefinestyle{eQASMStyle}{
  language=eQASM,
  linewidth=0.97\columnwidth,
  xleftmargin=0.06\columnwidth,
  basicstyle=\linespread{1.2}\small\ttfamily, 
  captionpos=b, 
  extendedchars=true, 
  tabsize=2, 
  columns=fixed, 
  keepspaces=true, 
  breaklines=true, 
  frame=trbl, 
  frameround=tttt, 
  numbers=left, 
  commentstyle=\color{eclipseGreen}, 
  keywordstyle=\color{eclipseGreen}, 
  keywordstyle=[2]*\bfseries\color{black}, 
  keywordstyle=[3]*\color{sublimegreen}, 
  keywordstyle=[4]\color{grayosix},
  keywordstyle=[5]*\bfseries\color{orange}, 
  keywordstyle=[6]*\color{eclipseBlue}, 
  keywordstyle=[7]\color{eclipsePurple}, 
  keywordstyle=[8]*\color{eclipseRed}
}

\lstnewenvironment{eQASM}%
{\lstset{style=eQASMStyle}}
{}

%% file: utils/behaviordescription.tex
\usepackage{listings}

\lstdefinelanguage{behavirodescription}
{
  sensitive=false, 
  keywords={},
  morekeywords=[2]{if, for, end, else, in, return, GPRF},
  morekeywords=[3]{register, MemUnit, integer, unsigned, bitstring, bool},
  morekeywords=[4]{UInt, SInt, MemWord, MemWord_val, MemByte, MemByte_val, GPR,  GPR_val, QOTRS, QOTRT, QMRR, QOTRS_val, QOTRT_val, QMRR_val, ToSBitStr, ToUBitStr, SignExt, ZeroExt, CompFlag_val},
  morekeywords=[5]{PC},
  morekeywords=[6]{Rs, Rt, Rd},
  morekeywords=[7]{assert},
  morekeywords=[8]{if, for, end, else, in, return, GPRF, QOTRFS, QOTRFT, QMRRF},
  morecomment=[l]{\#}, 
}

\lstdefinestyle{OpBehaviorStyle}{
  language=behavirodescription,
  linewidth=0.9\columnwidth,
  xleftmargin=0.1\columnwidth,
  basicstyle=\linespread{1.2}\footnotesize\ttfamily, 
  captionpos=b, 
  extendedchars=true, 
  tabsize=2, 
  columns=fixed, 
  keepspaces=true, 
  breaklines=true, 
  frame={tRBl}, 
  frameround=ffff, 
  numbers=none, 
  mathescape=true,
  literate=
     {=}{$\leftarrow{}$}{1}
     {==}{$={}={}$}{1}
     {<=}{$\le$}{1}
     {>=}{$\ge$}{1}
     {!=}{$\ne$}{1},
  commentstyle=\color{eclipseGreen}, 
  keywordstyle=\color{eclipseGreen}, 
  keywordstyle=[2]\bfseries\color{black}, 
  keywordstyle=[3]\color{sublimegreen}, 
  keywordstyle=[4]\color{eclipsePurple},
  keywordstyle=[5]*\bfseries\color{orange}, 
  keywordstyle=[6]\color{eclipseBlue},
  keywordstyle=[7]\color{eclipseRed}, 
  keywordstyle=[8]\bfseries
}

\lstnewenvironment{OpBehavior}
{\lstset{style=OpBehaviorStyle}}
{}

%% file: utils/insn_table_header.tex
\newcommand{\longtableheader}{ & & & & & & & & & & & & & & & & & & & & & & & & & & & & & & & \\
\multicolumn{1}{@{}c@{}}{\small31} & \multicolumn{1}{@{}c@{}}{\small30} & \multicolumn{1}{@{}c@{}}{\small29} & \multicolumn{1}{@{}c@{}}{\small28} & \multicolumn{1}{@{}c@{}}{\small27} & \multicolumn{1}{@{}c@{}}{\small26} & \multicolumn{1}{@{}c@{}}{\small25} & \multicolumn{1}{@{}c@{}}{\small24} & \multicolumn{1}{@{}c@{}}{\small23} & \multicolumn{1}{@{}c@{}}{\small22} & \multicolumn{1}{@{}c@{}}{\small21} & \multicolumn{1}{@{}c@{}}{\small20} & \multicolumn{1}{@{}c@{}}{\small19} & \multicolumn{1}{@{}c@{}}{\small18} & \multicolumn{1}{@{}c@{}}{\small17} & \multicolumn{1}{@{}c@{}}{\small16} & \multicolumn{1}{@{}c@{}}{\small15} & \multicolumn{1}{@{}c@{}}{\small14} & \multicolumn{1}{@{}c@{}}{\small13} & \multicolumn{1}{@{}c@{}}{\small12} & \multicolumn{1}{@{}c@{}}{\small11} & \multicolumn{1}{@{}c@{}}{\small10} & \multicolumn{1}{@{}c@{}}{\small9} & \multicolumn{1}{@{}c@{}}{\small8} & \multicolumn{1}{@{}c@{}}{\small7} & \multicolumn{1}{@{}c@{}}{\small6} & \multicolumn{1}{@{}c@{}}{\small5} & \multicolumn{1}{@{}c@{}}{\small4} & \multicolumn{1}{@{}c@{}}{\small3} & \multicolumn{1}{@{}c@{}}{\small2} & \multicolumn{1}{@{}c@{}}{\small1} & \multicolumn{1}{@{}c@{}}{\small0}\\
\cline{1-32}
\endhead
\cline{1-32}
\endfoot}

\newcommand{\qinsnheader}{ & & & & & & & & & & & & & \\
 \multicolumn{1}{@{}c@{}}{\small13} & \multicolumn{1}{@{}c@{}}{\small12} & \multicolumn{1}{@{}c@{}}{\small11} & \multicolumn{1}{@{}c@{}}{\small10} & \multicolumn{1}{@{}c@{}}{\small9} & \multicolumn{1}{@{}c@{}}{\small8} & \multicolumn{1}{@{}c@{}}{\small7} & \multicolumn{1}{@{}c@{}}{\small6} & \multicolumn{1}{@{}c@{}}{\small5} & \multicolumn{1}{@{}c@{}}{\small4} & \multicolumn{1}{@{}c@{}}{\small3} & \multicolumn{1}{@{}c@{}}{\small2} & \multicolumn{1}{@{}c@{}}{\small1} & \multicolumn{1}{@{}c@{}}{\small0}\\
\cline{1-14}
\endhead
\cline{1-14}
\endfoot}

\newenvironment{isatable}
{\vspace{-1cm}\begin{longtable}{*{32}{@{}p{4mm}}@{}}\longtableheader}
{\end{longtable}}

\newcommand{\reserved}{\textcolor{grayosix}{reserved}}

%% file: 00-intro.tex
\section{Introduction}
Executable QASM (eQASM) is an executable Quantum Instruction Set Architecture (QISA) for the quantum accelerator in a heterogeneous architecture as proposed in~\cite{fu2018eqasm}. eQASM contains both quantum instructions and auxiliary classical instructions. eQASM supports a set of \textit{discrete} quantum operations which can be defined via configuration before runtime. eQASM features accurate timing, Single-Operation-Multiple-Qubit (SOMQ) execution, VLIW architecture, programmable runtime feedback, and operational implementation.

We \textbf{instantiated} eQASM targeting a seven-qubit superconducting quantum processor. Since the binary of this eQASM instantiation can be executed by a microarchitecture implemented in the device QuTech Central Controller-Light (CC-Light), we call this \textbf{instantiation} \textit{CC-Light eQASM} to distinguish it from eQASM itself and other eQASM instantiation.

This document serves as a reference manual for both CC-Light users who write program at the QISA level and compiler developers who is going to develop backends for CC-Light eQASM instructions. Since the paper~\cite{fu2018eqasm} explains the eQASM architecture from a higher-level perspective, this document mostly focus on the specification of CC-Light eQASM to avoid redundancy.

This document is organized as following. An overview of the CC-Light eQASM is shown in Section~\ref{sec:overview}.
The formal eQASM assembly syntax is shown in Section~\ref{sec:syntax}. Section~\ref{sec:alphabet} presents the alphabet list of CC-Light eQASM instructions except the quantum bundle instructions. The appendix give some examples to illustrate how to use eQASM to perform quantum experiments and algorithms. Also, the format of the configuration file used to configure the assembler is given.

%% file: 01-overview.tex
\section{eQASM Overview}
\label{sec:overview}
\begin{figure}[hbt]
\centering
\includegraphics[width=0.8\columnwidth]{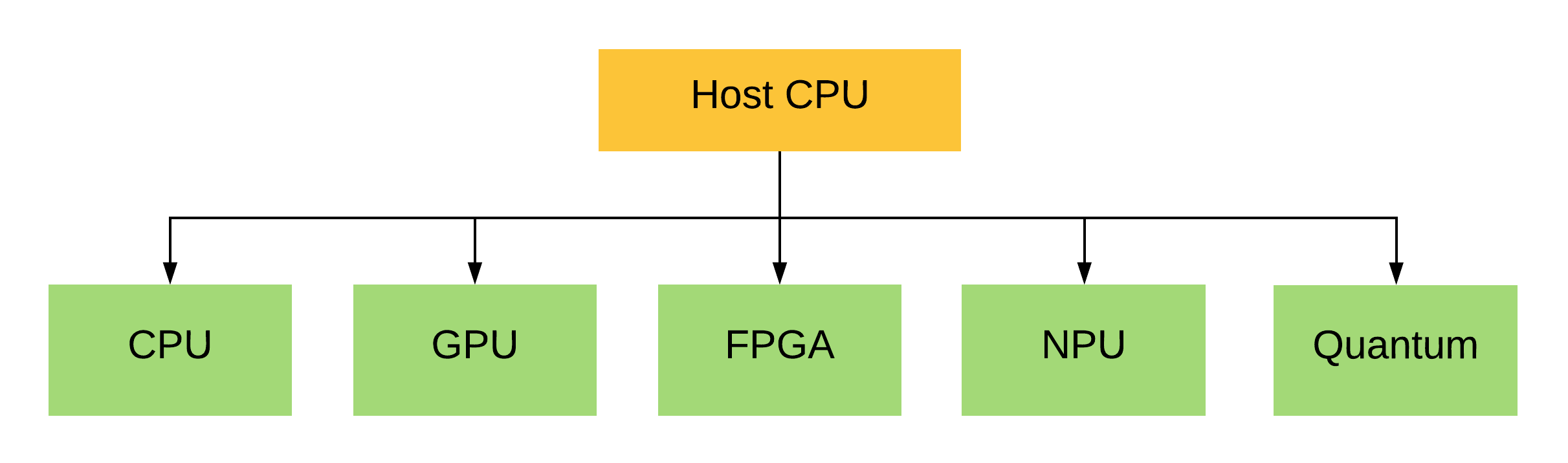}
\caption{Quantum processor as an accelerator in a heterogeneous architecture.}
\label{fig:heterogeneous}
\end{figure}

Similar to GPU or NPU, quantum computing can be integrated in a heterogeneous architecture as shown in Fig.~\ref{fig:heterogeneous}. The quantum part can be seen as a coprocessor used to accelerate particular tasks. eQASM only describes computation tasks executed on the quantum coprocessor.

This section introduces the eQASM programming and compilation model, the quantum program lifecycle, the architectural state, and an overview of instructions.

\subsection{Programming and Compilation Model}
\begin{figure}[bt]
\centering
\hspace{-1cm}
\includegraphics[width=0.5\columnwidth]{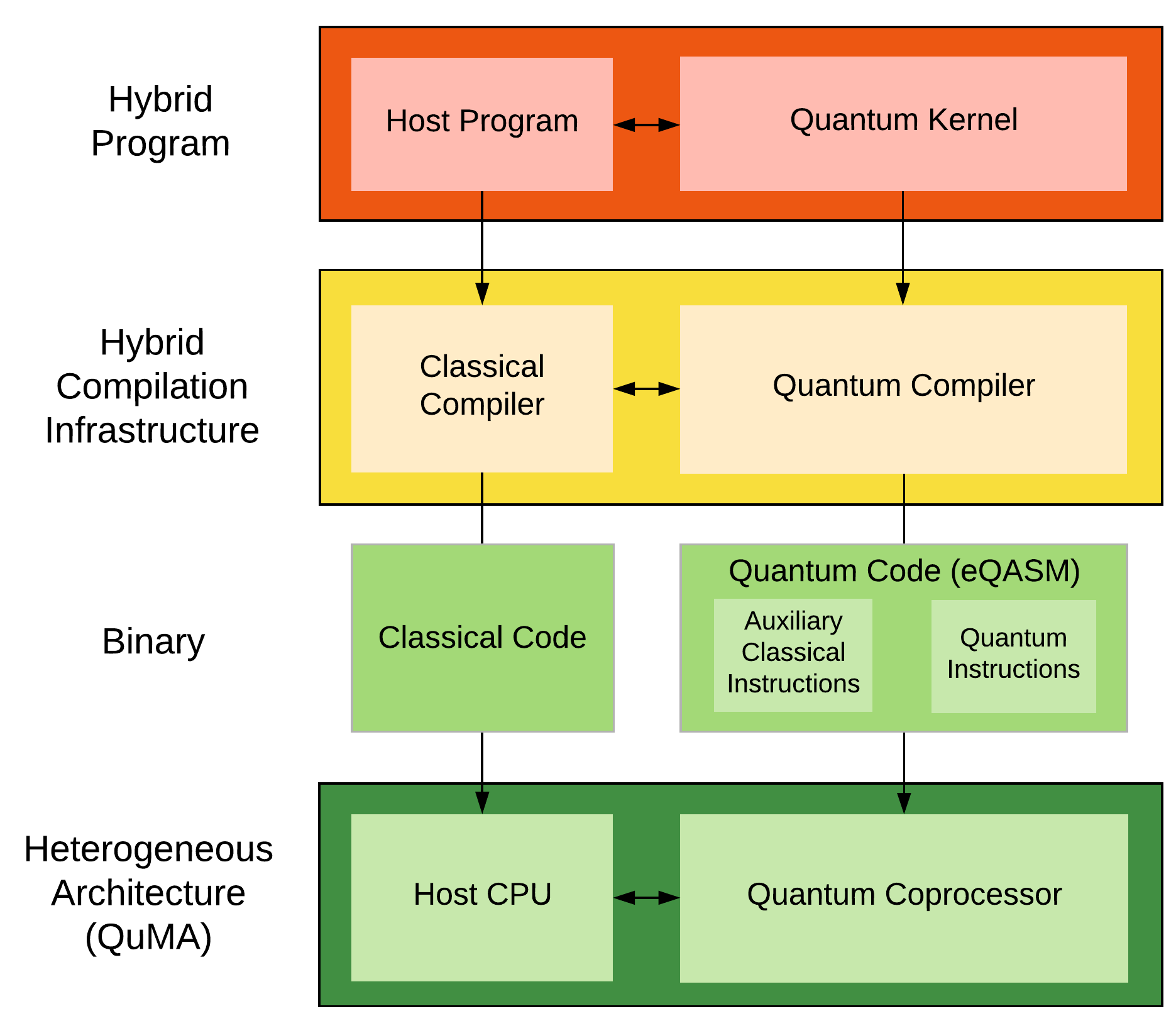}
\caption{Heterogeneous quantum programming and compilation model.}
\label{fig:prog_model}
\end{figure}

eQASM adopts the programming and compilation model as illustrated in Fig.~\ref{fig:prog_model}. A quantum-classical hybrid program contains a host program and one or more quantum kernels. During execution, the host program invokes the quantum kernel(s) to accelerate a particular part of the computation.
A hybrid compilation infrastructure compiles the hybrid program into classical code and quantum code which are fed to the heterogeneous architecture and directly executed by the host CPU and quantum coprocessor, respectively.

\subsection{Quantum Program Lifecycle}
A quantum program lifecycle using eQASM architecture contains the following phases: edit time, configuration time, compile time, and run time (or execution time).

\subsubsection{Edit Time}
In this phase, quantum programmers describe the quantum application or experiments using high-level languages. The host program can be described using a classical language, such as C++ or Python. The quantum kernel is described using a high-level quantum programming language, such as OpenQL. The host program should contain methods that offload the kernel to the accelerator for execution and read the result from the accelerator after the execution finishes. The current eQASM programming model does not define how these methods should be implemented.

\subsubsection{Configuration Time}
Since quantum instructions are not fully defined in eQASM, this phase is used to completely configure the instruction set and the hardware. This process contains the following aspects:
\begin{enumerate}
    \item \label{step:prim} Define available pulses that can be applied on qubits, with each pulse corresponding to a \textbf{primitive quantum operation} (for single-qubit gates) or part of a primitive quantum operation (for two-qubit gates). Upload all these pulses into the codeword-triggered pulse generator of the microarchitecture and assign an unique codeword to each pulse. The duration of each operation can be calculated at this step.
    \item \label{step:decomp} Define i) all available quantum instructions and ii) the decomposition of each quantum instruction to the primitive quantum operations as defined in step~(\ref{step:prim}). Each quantum instruction is assigned with an unique \textit{opcode}. The decomposition is described by a map from one opcode to one or multiple codewords with correct timing. The duration of each operation calculated in step~(\ref{step:prim}) is used to ensure correct timing.
    \item Based on the opcode and decomposition defined in step~(\ref{step:decomp}), configure the assembler to translate each quantum instruction into a correct opcode, and the microcode unit in the microarchitecture to perform the correct decomposition.
\end{enumerate}

\subsubsection{Compile Time}
In the hybrid compilation infrastructure, a conventional compiler, such as GNU Compilation Collection (GCC), compiles the host program into classical code. The quantum compiler, such as OpenQL~\cite{fu2017experimental}, compiles the quantum kernels into quantum code consisting of eQASM instructions, which contains quantum instructions.

\subsubsection{Run Time}
The host program is executed on the host CPU. When the program execution reaches particular points, the host CPU loads the quantum code of the desired kernel as well as the necessary initialization data into the quantum coprocessor where the quantum code can be directly executed. After the kernel execution finishes, the quantum coprocessor writes the result into a shared memory space where the host CPU can fetch the result for the following processing. The current eQASM design does define the detailed mechanism of how host CPU and the quantum processor interacts.

\subsection{Architectural State}
\label{sec:sys_state}
\begin{figure}[hbt]
\centering
\includegraphics[width=0.5\columnwidth]{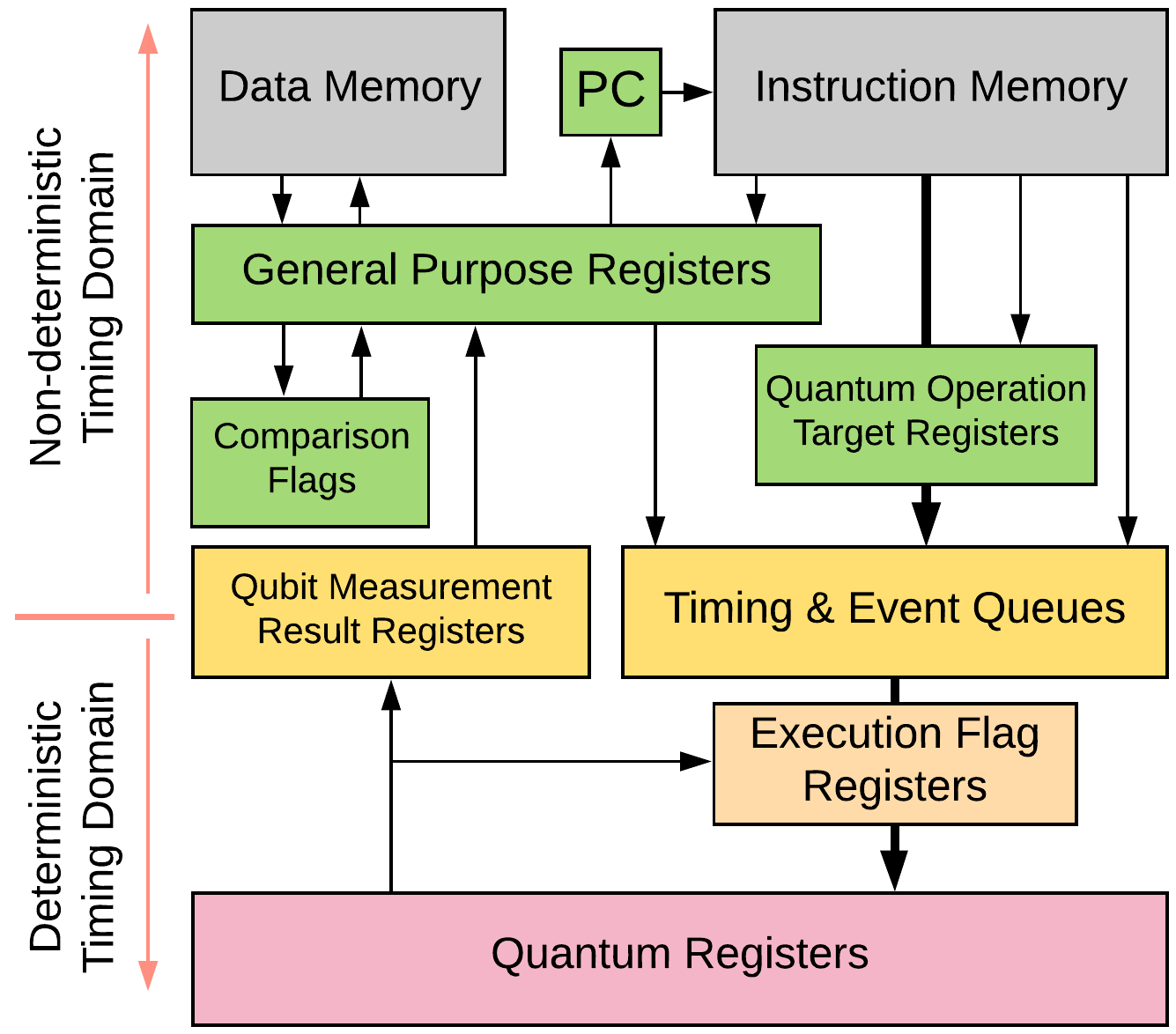}
\caption{Architectural state of eQASM. Arrows indicates the possible information flow. The thick arrows represent quantum operations, which reads information from the modules passed through. }
\label{fig:arch_state}
\end{figure}

As shown in Fig.~\ref{fig:arch_state}, the architectural state of the quantum coprocessor includes: a data memory, an instruction memory, a program counter (PC), a general purpose register (GPR) file, comparison flags, a quantum operation target register file, timing and event queues, a qubit measurement result register file, an execution flag register file, and a quantum register file.

\subsubsection{Data Memory}
The data memory can buffer intermediate computation results and serve as the communication channel between the host CPU and the quantum coprocessor.

\subsubsection{Instruction Memory \& Program Counter}
The eQASM instructions are stored in the instruction memory, and the Program Counter (PC) should contain the address of the next eQASM instruction to fetch.

\subsubsection{General Purpose Registers}
The general purpose register (GPR) file is a set of 32 registers, labeled as \code{Ri}, where $i\in\{0, 1, \ldots, 31\}$ is the register address. Each GPR is 32 bits wide.

\subsubsection{Comparison Flags}\label{sec:comp_flag_reg_file}
The comparison flags store the comparison result of two general purpose registers which are used by comparison (\lstinline!CMP!) and branch related instructions (\lstinline!BR!, \lstinline!FBR!). Table~\ref{tab:cmp_flags} lists the comparison flags with the corresponding meaning defined in eQASM. Function \code{unsigned(Ri, 32)} returns the 32-bit unsigned value stored in GPR \code{Ri}, and function \code{signed(Ri, 32)} returns the 32-bit signed value stored in GPR \code{Ri}. \lstinline!COMPFLAG! represents the collection of all comparison flags. Each flag can be accessed using \lstinline!COMPFLAG.<flag>! where \code{<flag>} is the name of the corresponding flag.
\begin{table}[hbt]
\centering
\caption{Comparison flags defined in eQASM.}
\label{tab:cmp_flags}
\small
\begin{tabular}{|c|c|c|c|}
\hline
\textbf{\textless Comparison Flag\textgreater} & \textbf{Meaning}                    & \textbf{\textless Comparison Flag\textgreater} & \textbf{Meaning}                    \\ \hline
ALWAYS                                     & 1                                   & NEVER                                      & 0                                   \\ \hline
EQ                                         & Rt == Rs                            & NE                                         & Rt != Rs                            \\ \hline
LTU                                        & unsigned(Rt) $<$ unsigned(Rs)       & GEU                                        & unsigned(Rt) $\ge$ unsigned(Rs)     \\ \hline
LEU                                        & unsigned(Rt) $\le$ unsigned(Rs)     & GTU                                        & unsigned(Rt) $>$ unsigned(Rs)       \\ \hline
LT                                         & signed(Rt, 32) $<$ signed(Rs, 32)   & GE                                         & signed(Rt, 32) $\ge$ signed(Rs, 32) \\ \hline
LE                                         & signed(Rt, 32) $\le$ signed(Rs, 32) & GT                                         & signed(Rt, 32) $>$ signed(Rs, 32)   \\ \hline
\end{tabular}
\end{table}

\subsubsection{Quantum Operation Target Registers}
Each quantum operation target register can be used as an operand of a quantum operation. There are two types of quantum operation target registers: 32 single-qubit target registers for single-qubit operations, and 32 two-qubit target registers for two-qubit operations. A single- (two-)qubit target register is labelled as \code{Si} (\code{Ti}), where $i\in\{0, 1, \ldots, 31\}$ is the register address.

\subsubsection{Timing and Event Queues}
eQASM adopts a queue-based timing control scheme. The timing and event queues are used to buffer timing points and quantum operations generated from the execution of quantum instructions (see Section~III-A of \cite{fu2018eqasm}).

\subsubsection{Qubit Measurement Result Registers}
Each qubit is associated with a qubit measurement result register (QMRR) with each being 1 bit wide. Each QMRR stores the result of the last finished measurement instruction on the corresponding qubit when it is valid. It is labeled as \code{Qi}, where $i\in\{0, 1, \ldots, 6\}$ is the physical address of the qubit.

\subsubsection{Execution Flag Registers}
\label{sec:exe_flag_reg}
The execution flag register file contains an execution flag register for each qubit to support fast conditional execution. Each execution flag register contains four execution flags of which the values are derived automatically by the microarchitecture from the last measurement results of the corresponding qubit.

eQASM uses the following combinatorial logic to define each execution flag:
\begin{enumerate}
    \item \bin{1} (the default for unconditional execution);
    \item \bin{1} if and only if (\textbf{iff}) the last finished measurement result is $\ket{1}$;
    \item \bin{1} \textbf{iff} the last finished measurement result is $\ket{0}$;
    \item \bin{1} \textbf{iff} the last two finished measurements get the same result.
\end{enumerate}
Note, the last finished measurement result refers to the result of the last finished measurement instruction on this qubit when these flags are used. It is irrelevant to the validity of the quantum measurement result register.

\subsubsection{Quantum Register}
The quantum register file is the collection of all seven physical qubits inside the quantum coprocessor. Each qubit is assigned a unique index, known as the \textit{physical address}. Since data in qubits can be superposed, eQASM does \textbf{not} allow direct access to the data at the instruction level. Instead, users can measure qubits using measurement instructions and later access the results in the qubit measurement result registers.

\subsection{Instruction Overview}

\begin{table}[bt]
\centering
\caption{Overview of eQASM Instructions.}
\label{tab:eqasm_insn}
\footnotesize
{\setlength{\extrarowheight}{2pt}%
\begin{tabular}{|c|l|l|}
\hline
\textbf{Type}      & \textbf{Syntax} & \textbf{Description} \\ \hline \hline
\multirow{2}{*}{Control}            & \lstinline!CMP  Rs, Rt! & \begin{tabular}[c]{@{}l@{}}Compare GPR \texttt{Rs} and \texttt{Rt} and store the result into the
                                                                        comparison flags.\end{tabular} \\ \cline{2-3}
                                    & \lstinline!BR   <Comp. Flag>, Offset! & \begin{tabular}[c]{@{}l@{}}Jump to \texttt{PC~+~Offset} if the specified comparison flag is \bin{1}.\end{tabular} \\ \hline
\multirow{6}{*}{Data Transfer} & \lstinline!FBR  <Comp. Flag>, Rd! & Fetch the specified comparison flag into GPR Rd. \\ \cline{2-3}
                               & \lstinline!LDI  Rd, Imm!   & Rd = signed\_ext(Imm{[}19..0{]}, 32). \\ \cline{2-3}
                               & \lstinline!LDUI Rd, Imm, Rs! & Rd = Imm{[}14..0{]}::Rs{[}16..0{]}. \\ \cline{2-3}
                               & \lstinline!LD   Rd, Rt(Imm)! & Load data from memory address \texttt{Rt + Imm} into GPR \texttt{Rd}.  \\ \cline{2-3}
                               & \lstinline!ST   Rs, Rt(Imm)! & Store the value of GPR \texttt{Rs} in memory address \texttt{Rt + Imm}. \\ \cline{2-3}
                               & \lstinline!FMR  Rd, Qi! & \begin{tabular}[c]{@{}l@{}}Fetch the result of the last measurement instruction on qubit \texttt{i} into GPR \texttt{Rd}.\end{tabular}\\ \hline
Logical & \begin{tabular}[c]{@{}l@{}}\lstinline!AND/OR/XOR Rd, Rs, Rt!\\ \lstinline!NOT  Rd, Rt!\end{tabular} & Logical and, or, exclusive or, not. \\ \hline
Arithmetic                     & \lstinline!ADD/SUB Rd, Rs, Rt! & Addition and subtraction. \\ \hline
\hline
Waiting & \begin{tabular}[c]{@{}l@{}} \lstinline!QWAIT   Imm!\\ \lstinline!QWAITR  Rs! \end{tabular} & \begin{tabular}[c]{@{}l@{}} Specify a timing point by waiting for the number of cycles indicated by \\ the immediate value \texttt{Imm} or the value of GPR \texttt{Rs}.\end{tabular} \\ \hline
Target Specify & \begin{tabular}[c]{@{}l@{}} \lstinline!SMIS Sd, <Qubit List>! \\ \lstinline!SMIT Td, <Qubit Pair List>! \end{tabular} & Update the single- (two-)qubit operation target register \texttt{Sd} (\texttt{Td}).\\ \hline
Q. Bundle &   \lstinline![PI,] Q_Op [| Q_Op]*! & \begin{tabular}[c]{@{}l@{}}Applying operations on qubits after waiting for a small number of \\cycles indicated by \texttt{PI}.\end{tabular} \\ \hline
\end{tabular}}
\end{table}

An eQASM program can consist of quantum instructions and auxiliary classical instructions, which can be interleaved in the instruction memory. An overview of eQASM instructions is shown in Table~\ref{tab:eqasm_insn}.
The top part of Table~\ref{tab:eqasm_insn} contains the auxiliary classical instructions. There are four types: \textit{control}, \textit{data transfer}, \textit{logical} and \textit{arithmetic} instructions. These are all scalar instructions.  The function \code{sign\_ext(Imm, 32)} sign-extends the immediate value \code{Imm} to \bits{32}. The operator \code{::} concatenates the two bit strings.

The bottom part of Table~\ref{tab:eqasm_insn} contains the quantum instructions. There are three types of instructions:
\begin{itemize}
    \item Explicit waiting instructions used to specify timing points (\lstinline!QWAIT!, \lstinline!QWAITR!),
    \item The quantum operation target register setting instructions (\lstinline!SMIS!, \lstinline!SMIT!), and
    \item The quantum bundle instructions, which consist of the specification of a small waiting time and multiple quantum operations.
\end{itemize}

The next section shows the formal eQASM assembly syntax.

%% file: 02-eqasm_syntax.tex
\section{eQASM Assembly Syntax Specification}
\label{sec:syntax}
This section describes the eQASM assembly syntax.

\lstset{style=eQASMStyle}
\subsection{File Organization}
A single eQASM program should be written in a single assembly file which contains a sequence of lines. The following rules applies:
\begin{itemize} \tightlist
    \item Each line ends with a newline character (ASCII CR+LF).
    \item All characters are case \textbf{in}sensitive, and \textit{extra} blank is allowed between two identifiers.
    \item \textbf{User-defined identifiers} can be used for mnemonic representations with the following rules:
    \begin{itemize}
        \item They should start with a letter or an underscore, and followed by letters, digits, or underscores; and
        \item They must not be a register name or an instruction name.
    \end{itemize}
    \item \textbf{Immediate values} can be specified in three different formats:
    \begin{itemize}\tightlist
        \item \textit{Plain format} has no prefix and is interpreted as base-10 numbers. E.g., \code{23}.
        \item \textit{Hexadecimal format} starts with the prefix \code{0x}. E.g., \code{0x17}.
        \item \textit{Binary format} starts with the prefix \code{0b}. E.g., \code{0b10111}.
    \end{itemize}
    \item Only \textbf{line comment} is supported.
    \begin{itemize}
        \item It start with a hash mark (\emph{\#}) and continues to the end of the line.
    \end{itemize}
    \item A line can be an \textbf{empty line}, or a \textbf{statement line}.
    \begin{itemize} \tightlist
        \item An empty line is a line with only white space, which can be spaces, tabs, and/or comments.
        \item The syntax of a statement line is defined in Section~\ref{instruction-line}.
    \end{itemize}
\end{itemize}

\hypertarget{instruction-line}{%
\subsection{Statement}\label{instruction-line}}
A statement line can contain one of the three kinds of statements:
\begin{itemize}
  \item \textbf{A directive statement}: a directive to the assembler that does not necessarily generate any code (see Section~\ref{sec:directive});
  \item \textbf{A label statement}: a mnemonic representation of the address of the first instruction following the label (see Section~\ref{sec:label-statement}); and/or
  \item \textbf{A machine operation statement}: a single-format instruction or a quantum bundle (see Section~\ref{sec:machine_op_statement}) that generates one or multiple 32-bit instruction words.
\end{itemize}
Except that a label statement can be followed by a machine operation statement in the same line, not any two statements can be put in the same line.

\hypertarget{assembler-directives}{%
\subsection{Directive Statement}\label{sec:directive}}
A directive can be a \hyperlink{register-aliases}{register alias} or a \hyperlink{constant-alias}{constant alias}. It is advised to put the directives at the top of a program.

\hypertarget{register-aliases}{%
\subsubsection{Register aliases}\label{register-aliases}}

Architecture registers can be given more meaningful names with the \lstinline!.register! keyword:
\begin{eQASM}
.register <Register Name> <Alias Name>
\end{eQASM}
where \code{\textless Alias Name\textgreater} should be a valid user-defined identifiers.

\paragraph{Example} The following code gives \code{s7} the alias \lstinline!all_qubits!, which indicates that \code{s7} is going to be used to contain all seven qubits.
\lstset{language=eQASM}
\begin{eQASM}
.register  s7 all_qubits
SMIS       s7, {0, 1, 2, 3, 4, 5, 6}
H          all_qubits
\end{eQASM}

\hypertarget{constant-alias}{%
\subsubsection{Constant Alias}\label{sec:constant-alias}}
Numerical constants can be given a more meaningful name with the \lstinline!.def_sym! keyword:
\lstset{language=eQASM}
\begin{eQASM}
.def_sym <Alias Name> <Immediate>
\end{eQASM}
where \code{\textless Alias Name\textgreater} should be a valid user-defined identifiers.
\paragraph{Example} The following code gives the constant 10000 the alias \lstinline!INIT_INTERVAL!, which indicates an interval of \us{200} when used by the \lstinline!QWAIT! instruction.
\begin{eQASM}
.def_sym INIT_INTERVAL 10000
QWAIT INIT_INTERVAL
\end{eQASM}

\hypertarget{sec:label-statement}{%
\subsection{Label Statement}\label{sec:label-statement}}
A label statement is a mnemonic representation of the address of the first instruction following the label. Labels can be used as the target of branch/jump instructions so that the programmer does not need to know the actual address of the target instruction.

Except blank, a label statement line starts with a label name followed by a colon. The label name should be a valid user-defined identifier. Every label should be unique across the eQASM program. A machine operation statement is allowed to be appended to a label statement in the same line.

Example 1:
\begin{eQASM}
Label: bs 1 X s0   # a label statement followed by a machine operation
                   #   statement
    ...            # some other code here
    goto Label     # jump back
\end{eQASM}

Example 2:
\begin{eQASM}
_AnotherValidLabel:          # a label statement
    bs 1 X s0                # a machine operation statement
    ...                      # some other code here
    goto _AnotherValidLabel  # jump back
\end{eQASM}

Example 1 and Example 2 are equivalent. The label \code{Label} or \code{\_AnotherValidLabel} represents the address of the instruction \code{bs 1 X s0}. This label is then used by the \code{goto} instruction to form an infinite loop.

\subsection{Machine Operation Statement}
\label{sec:machine_op_statement}
A machine operation statement can be a single-format instruction or a quantum bundle. A single instruction contains all auxiliary classical instructions, and the \lstinline!QWAIT!, \lstinline!QWAITR!, \lstinline!SMIS!, or \lstinline!SMIT! instruction, as shown in Table~\ref{tab:eqasm_insn}.
Every single-format instruction is encoded into a single instruction word (32-bit).
The syntax, encoding, and behavior of each instruction is presented in the alphabet list of instructions in Section~\ref{sec:alphabet}.

The assembly format of a quantum bundle is defined as following with \code{PI} ranging from 0 to 7:
\begin{center}
\lstinline![PI,]  <Quantum Operation>  [| <Quantum Operation>]*!
\end{center}

A quantum operation can be one of the three types as shown in table~\ref{tab:qop_syntax}.

\begin{table}[hbt]
    \centering
    \begin{tabular}{|c|l|}
    \hline
    \textbf{Quantum Operation Type}     &  \textbf{Syntax}\\ \hline
    No Operation    & \lstinline!QNOP! \\ \hline
    Single-qubit    & \lstinline!<single_qubit_operation_name> Si!\\ \hline
    Two-qubit       & \lstinline!<two_qubit_operation_name>    Ti! \\ \hline
    \end{tabular}
    \caption{Quantum operation types and the syntax.}
    \label{tab:qop_syntax}
\end{table}

A quantum bundle might be translated into one or multiple 32-bit \textit{quantum bundle instructions} depending on if the number of quantum operations in the bundle is larger than 2 or not. The binary format of a quantum bundle instruction is shown in Table~\ref{tab:bundle-format}. The most significant bit (MSb) being \bin{1} means that this is a quantum bundle instruction instead of a single-format instruction whose MSb is \bin{0}.
\begin{table}[H]
\centering
\small
\begin{isatable}
\multicolumn{ 1}{|@{}c@{} }{\small          1} &
\multicolumn{ 9}{|@{}c@{} }{\small q\_opcode\_0} &
\multicolumn{ 5}{|@{}c@{} }{\small Si/Ti\_0} &
\multicolumn{ 9}{|@{}c@{} }{\small q\_opcode\_1} &
\multicolumn{ 5}{|@{}c@{} }{\small Si/Ti\_1} &
\multicolumn{ 3}{|@{}c@{}|}{\small PI}
\\
\cline{1-32}
\end{isatable}
\caption{Quantum bundle format.}
\label{tab:bundle-format}
\end{table}

The single-qubit and two-qubit operation names and their opcodes are user-definable, which are specified using a configuration file \filename{qisa\_opcodes.qmap}. The file \filename{qisa\_opcodes.qmap} will be read by the assembler. The specification of \filename{qisa\_opcodes.qmap} is shown in Appendix~\ref{sec:qmap}. The control store of the microcode unit (see~\cite{fu2018eqasm} for more details) should also be configured in a consistent way. The specification of the control store file used by the CC-Light driver and an example are shown in Appendix~\ref{sec:control_store}. Since the quantum opcode width is 9, the user can define at most \textbf{511} quantum operations except the \lstinline[style=eQASMstyle]!QNOP! operation during one run of the quantum operation.

\subsection{Predefined Macros}

The CC-Light assembler also includes some predefined MACROs to enable easy programming. For example, the instruction \lstinline!BLTU Rs,Rt, tgt_addr! will be decomposed by the assembler to two separate instructions:
\begin{center}
\lstinline[style=eQASMstyle]!CMP Rs, Rt!\\
\lstinline[style=eQASMstyle]!BR LTU, tgt_addr!
\end{center}
It first compares \code{Rs} and \code{Rt}, and changes the PC to the target address if \code{unsigned(Rs)} is less than \code{unsigned(Rt)}.

Table~\ref{tab:macro} shows the predefined macros in CC-Light eQASM. Note, if one macro is decomposed into two instructions, then these two instructions will be put into two consecutive lines by the assembler.

\begin{table}[H]
    \centering
    \caption{Predefined macros in CC-Light eQASM.}
    \label{tab:macro}
    \small
    \begin{tabular}{|l|l|l|}
        \hline
        Macro & Intepretation & Comment\\ \hline
        \lstinline!GOTO  addr!           & \lstinline!BR always, addr!                          & PC $\leftarrow$ addr~$\ll$~2 \\ \hline
        \lstinline!BRN   addr!            & \lstinline!BR never,  addr!                          & No operation \\ \hline
        \lstinline!BEQ   Rs, Rt, addr!   & \lstinline!CMP Rs, Rt!\quad\qquad \lstinline!BR eq,  addr!      &  \\ \hline
        \lstinline!BNE   Rs, Rt, addr!   & \lstinline!CMP Rs, Rt!\quad\qquad \lstinline!BR ne,  addr!      &  \\ \hline
        \lstinline!BLT   Rs, Rt, addr!   & \lstinline!CMP Rs, Rt!\quad\qquad \lstinline!BR lt,  addr!      &  \\ \hline
        \lstinline!BLE   Rs, Rt, addr!   & \lstinline!CMP Rs, Rt!\quad\qquad \lstinline!BR le,  addr!      &  \\ \hline
        \lstinline!BGT   Rs, Rt, addr!   & \lstinline!CMP Rs, Rt!\quad\qquad \lstinline!BR gt,  addr!      &  \\ \hline
        \lstinline!BGE   Rs, Rt, addr!   & \lstinline!CMP Rs, Rt!\quad\qquad \lstinline!BR ge,  addr!      &  \\ \hline
        \lstinline!BLTU  Rs, Rt, addr!   & \lstinline!CMP Rs, Rt!\quad\qquad \lstinline!BR ltu, addr!     &  \\ \hline
        \lstinline!BLEU  Rs, Rt, addr!   & \lstinline!CMP Rs, Rt!\quad\qquad \lstinline!BR leu, addr!     &  \\ \hline
        \lstinline!BGTU  Rs, Rt, addr!   & \lstinline!CMP Rs, Rt!\quad\qquad \lstinline!BR gtu, addr!     &  \\ \hline
        \lstinline!BGEU  Rs, Rt, addr!   & \lstinline!CMP Rs, Rt!\quad\qquad \lstinline!BR geu, addr!     &  \\ \hline
        \lstinline!MOV   Rd, Rs!         & \lstinline!LDI Rd, 0 !\quad\qquad \lstinline!ADD Rd, Rs, Rd!   & Rd $\leftarrow$ Rs \\ \hline
        \lstinline!SHL1  Rd, Rs!         & \lstinline!ADD Rd, Rs, Rs!                           & Rd $\leftarrow$ Rs~$\ll$~1  \\ \hline
        \lstinline!MULT2 Rd, Rs!        & \lstinline!ADD Rd, Rs, Rs!                           & Rd $\leftarrow$ Rs~$\times$~2 \\ \hline
        \lstinline!NAND  Rd, Rs, Rt!     & \lstinline!AND Rd, Rs, Rt! \quad \lstinline!NOT Rd, Rd! &  \\ \hline
        \lstinline!NOR   Rd, Rs, Rt!      & \lstinline!OR  Rd, Rs, Rt! \quad \lstinline!NOT Rd, Rd! &  \\ \hline
        \lstinline!XNOR  Rd, Rs, Rt!     & \lstinline!XOR Rd, Rs, Rt! \quad \lstinline!NOT Rd, Rd! &  \\ \hline
    \end{tabular}
\end{table}

\subsection{Latency Between Instructions}
\label{sec:latency}
It is a common case that a later instruction $k_{j}$ maybe depend on a previous instruction  $k_{i}$, i.e., the instruction $k_j$ needs to read the register which was previously written by $k_i$. This is called data dependency.

QuMA\_v2 is implemented in a pipelined fashion, and not all data dependency is resolved by the hardware. Instead, the compiler should be aware of this fact and insert a small number of independent instructions between two inter-dependent instruction to make sure the later one is reading the correct value. We call this process \textit{instruction latency compensation}.

Two kinds of latency requires compensation: writing and reading comparison flags, writing and reading the measurement result (FMR).
NOTE, any writing and reading the same GPRs are automatically handled by the microarchitecture, which requires no compensation. Also, No branch slot is reserved in the CC-Light eQASM, in other words, all instructions following a \lstinline!BR! instruction will \textbf{not} be executed once the branch takes place.

\subsubsection{Comparison Flags}
One extra instruction is required between any instruction writing the comparison flags (\lstinline!cmp!) and any instruction reading the comparison flags (\lstinline!fbr! and \lstinline!br!).

\subsubsection{FMR}
To enable the comprehensive feedback control work properly, two independent instructions should be inserted between \textbf{any} measurement instructions and the following \lstinline!FMR! instruction, as shown in Fig.~\ref{fig:cfc}.

\begin{figure}[hbt]
    \begin{eQASM}
      SMIS  S0, {0}
      SMIS  S1, {1}
      LDI   R0, 1
      MeasZ S1
      QWAIT 30
      NOP               # two insns compensate for latency of MSMT -> FMR
      FMR   R1, Q1      # fetch msmt result
      CMP   R1, R0      # compare
      NOP               # one insn compensate for latency of CMP -> BR
      BR    EQ, eq_path # jump if R0 == R1
    ne_path:
      X     S0          # happen if msmt result is 0
      BR    ALWAYS, continue
    eq_path:
      Y     S0          # happen if msmt result is 1
    continue:
      ...
    \end{eQASM}
    \caption{Conditionally performing a gate on qubit 0 based on the measurement result of qubit 1 using comprehensive feedback control. \lstinline!NOP!s are inserted between dependent instructions.}
    \label{fig:cfc}
\end{figure}

%% file: 03-isa.tex
\section{Alphabet List of Predefined eQASM Instructions}
\label{sec:alphabet}
In this section, We start by defining some basic functions using pseudo code, which are used in describing the behavior of instructions. The alphabet list of the predefined eQASM instructions are given in the following subsection.

\subsection{Numbering \& Operators}
In CC-Light eQASM, binary data is Most Significant bit (MSb) first (i.e. the MSb is on the left side of the bit string) and ``LSb 0'' bit numbering (i.e., the Least Significant bit (LSb) is numbered as 0). For example, the constant \code{0x5} is represented as \code{0b00000101} in a eight-bit binary representation, and bit 0 is of the value \bin{1}.

Table~\ref{tab:operator} defines the operators used in the instruction description.
\begin{table}[hbt]
\centering
    \caption{Operators with the meaning and precedence used in this manual. Note, a smaller number indicates a higher precedence.}
    \begin{tabular}{|c|c|c|}
    \hline
    \textbf{Operator} & \textbf{Meaning} & Precedence\\ \hline
      $\code{bitstr}\left[\code{h} : \code{l}\right]$ & \begin{tabular}[l]{@{}l@{}}Slice the bit string \code{bitstr}, with the range from the \code{h}-th bit down to the \code{l}-th bit. \end{tabular} & 0 \\ \hline
      $\left[\code{start} .. \code{incr} .. \code{end}\right]$ & \begin{tabular}[l]{@{}l@{}}Generate an iterable list with a step of \code{incr} which starts from \code{start} and ends no \\greater than \code{end}.\end{tabular} & 0 \\ \hline
      ** & Exponent - left operand raised to the power of right & 1  \\ \hline
      \textasciitilde & Bitwise NOT the operand & 2 \\ \hline
      * & Multiply two operands & 3  \\ \hline
      / & Divide left operand by the right one & 3    \\ \hline
      \% & Modulus - remainder of the division of left operand by the right & 3\\ \hline
      + & Add two operands or unary plus  & 4      \\ \hline
      - & Subtract right operand from the left or unary minus & 4   \\ \hline
      $\ll$ & \begin{tabular}[l]{@{}l@{}} Left shift the left operand by the number of bits specified by the right operand\end{tabular} & 5 \\ \hline
      $\ge,~ >,~ \le,~ <$ & Relational operators & 6 \\ \hline
      $==, !=$ & Relational operators & 7 \\ \hline
      \& & Bitwise AND two operands & 8   \\ \hline
      $\wedge$ & Bitwise XOR two operands & 9   \\ \hline
      $|$ & Bitwise OR two operands & 10   \\ \hline
      $\leftarrow$ & assignment & 11 \\ \hline
    \end{tabular}
    \label{tab:operator}
\end{table}

\subsection{Functions}

\subsubsection{Memory, Registers and Comparison Flags}
Function \code{MemByte(bitstring<32> x)} returns the memory unit (MemUnit) with 8 bits whose address specified by the bit string \code{x}, and function \code{MemByte\_val(bitstring<32> x)} further returns the 8-bit value stored in the byte structure.
Function \code{MemWord(bitstring<32> x)} returns the word structure with the starting address specified by the bit string \code{x}, and function \code{MemWord\_val(bitstring<32> x)} further returns the 32-bit value stored in the word structure.
\begin{OpBehavior}
MemUnit<1> MemByte(bitstring<32> x):
    return Mem[unsigned_integer(x, 32)]

bitstring MemByte_val(bitstring<32> x):
    return bitstring<8>(MemByte(x))
    
MemUnit<4> MemWord(bitstring<32> x):
    return Mem[unsigned_integer(x, 32) + 3 : unsigned_integer(x, 32)]

bitstring MemWord_val(bitstring<32> x):
    return bitstring<32>(MemWord(x))
\end{OpBehavior}

Function \code{GPR(bitstring<5> x)} returns the general purpose register with the register number specified by the bit string \code{x}, and function \code{GPR\_val(bitstring<5> x)} further returns the 32-bit value stored in the register.
\begin{OpBehavior}
register GPR(bitstring<5> x):
    return GPRF[unsigned_integer(x, 5)]

bitstring GPR_val(bitstring<5> x):
    return bitstring<32>(GPR(x))
\end{OpBehavior}

Function \code{QOTRS(bitstring<5> x)} (\code{QOTRT(bitstring<5> x)}) returns the single- (two-)qubit quantum operation target register with the register number specified by the bit string \code{x}, and function \code{QOTRS\_val(bitstring<5> x)} (\code{QOTRT\_val(bitstring<5> x)}) further returns the 7- (16-)bit value stored in the register.
\begin{OpBehavior}
register QOTRS(bitstring<5> x):
    return QOTRFS[unsigned_integer(x, 5)]
    
bitstring QOTRS_val(bitstring<5> x):
    return bitstring<7>(QOTRS(x))
    
register QOTRT(bitstring<5> x):
    return QOTRFT[unsigned_integer(x, 5)]
    
bitstring QOTRT_val(bitstring<5> x):
    return bitstring<16>(QOTRT(x))
\end{OpBehavior}

Function \code{QMRR(bitstring<3> x)} returns the quantum measurement result register with the register number specified by the bit string \code{x}, and function \code{QMRR\_val(bitstring<5> x)} further returns the 1-bit value stored in the register.
\begin{OpBehavior}
register QMRR(bitstring<5> x):
    return QMRRF[unsigned_integer(x, 3)]
    
bitstring QMRR_val(bitstring<5> x):
    return bitstring<1>(QMRR(x))
\end{OpBehavior}

Function \code{CompFlag\_val} returns the value of the comparison flag \code{comp\_flag}. All comparison flags in CC-Light eQASM is listed in Table~\ref{tab:cmp_flags}.
\begin{OpBehavior}
bitstring CompFlag_val(comp_flag):
    return bitstring<1>(COMPFLAG.comp_flag)
\end{OpBehavior}

\subsubsection{UInt (Unsigned Int)}
This function returns the unsigned value of the least significant \code{N} bits in the bit string \code{x}.

\begin{OpBehavior}
integer UInt(bitstring<M> x, integer N):
    assert(M >= N)
    
    integer result = 0
    
    for i in [0 .. 1 .. (N - 1)]:
        if x[i] == '1':
            result = result + 2 ** i
        end if
    end for
            
    return result
\end{OpBehavior}

\subsubsection{SInt (Signed Int)}
This function returns the signed value of the least significant \code{N} bits in the bit string \code{x}.
\begin{OpBehavior}
integer SInt(bitstring<M> x, integer N):
    assert(M >= N)
    
    integer result = 0
    
    for i in [0 .. 1 .. (N - 2)]:
        if x[i] == '1':
            result = result + 2 ** i
        end if
    end for
    
    if x[N - 1] == '1':
        result = result - 2 ** (N - 1)
    end if
            
    return result
\end{OpBehavior}

\subsubsection{ToUBitStr (Convert Unsigned Integer to Bit String)}
This function returns the \code{N}-bit binary representation of the given unsigned integer \code{int\_val}.
\begin{OpBehavior}
bitstring ToUBitStr(integer int_val, integer N):
    assert(0 <= int_val <= 2 ** N - 1)
    
    bitstring<N> result = 0
    
    if int_val > 2 ** (N -1):
        result[N - 1] = 1;
        int_val = int_val - 2 ** (N - 1)
    end if
    
    for i in [0 .. 1 .. (N - 2)]:
        result[i] = int_val 
        int_val = (int_val - result[i]) / 2
    end for
    
    return result
\end{OpBehavior}

\subsubsection{ToSBitStr (Convert Signed Integer to Bit String)}
This function returns the \code{N}-bit 2's complement of the given signed integer \code{int\_val}.
\begin{OpBehavior}
bitstring ToSBitStr(integer int_val, integer N):
    assert(-2 ** (N - 1) <= int_val <= 2 ** (N - 1) - 1)
    
    bitstring<N> result = 0
    
    if signed:
        if int_val < 0:
            result[N - 1] = 1;
            int_val = int_val + 2 ** N
        end if
         
    for i in [0 .. 1 .. (N - 2)]:
        result[i] = int_val 
        int_val = (int_val - result[i]) / 2
    end for
    
    return result
\end{OpBehavior}

\subsubsection{ZeroExt (Unsigned Extend)}
This function unsigned-extends the given bitstring \code{x} to the given length \code{N}.
\begin{OpBehavior}
bitstring ZeroExt(bitstring<M> x, integer N):
    assert(M <= N)
    
    bitstring<N> result = 0
    
    result[0 : M - 1] = x[0 : M - 1]
    
    for i in [M, M + 1, ..., N - 1]:
        result[i] = 0
    end for
    
    return result
\end{OpBehavior}

\subsubsection{SignExt (Signed Extend)}
This function signed-extends the given bit string \code{x} to the given length \code{N}.
\begin{OpBehavior}
bitstring SignExt(bitstring<M> x, integer N):
    assert(M <= N)
    
    bitstring<N> result = 0
    
    result[0 : M - 1] = x[0 : M - 1]
    
    for i in [M .. 1 .. (N - 1)]:
        result[i] = x[M - 1]
    end for
    
    return result
\end{OpBehavior}

\input{isa/insn_encoding.tex}

%% file: isa/insn_encoding.tex
\subsection{ ADD -- Add }\label{sec:insn-add}
\begin{table}[H]
\centering
\begin{isatable}
\multicolumn{ 1}{|@{}c@{} }{\small          0} & 
\multicolumn{ 1}{|@{}c@{} }{\small          0} & 
\multicolumn{ 1}{ @{}c@{} }{\small          1} &
\multicolumn{ 1}{ @{}c@{} }{\small          1} &
\multicolumn{ 1}{ @{}c@{} }{\small          1} &
\multicolumn{ 1}{ @{}c@{} }{\small          1} &
\multicolumn{ 1}{ @{}c@{} }{\small          0} &
\multicolumn{ 5}{|@{}c@{} }{\small         Rd} & 
\multicolumn{ 5}{|@{}c@{} }{\small         Rs} & 
\multicolumn{ 5}{|@{}c@{} }{\small         Rt} & 
\multicolumn{10}{|@{}c@{}|}{\small  \reserved}   
\\
\cline{1-32}
\end{isatable}
\end{table}
\vspace{-0.4cm}

\quad\textbf{Format}: \hspace{4cm} \lstinline[basicstyle=\normalsize\ttfamily]!ADD Rd, Rs, Rt!

\textbf{Description:}
\begin{adjustwidth}{1cm}{1cm}
The \lstinline!ADD! instruction adds two GPR (\code{Rs}, \code{Rt}) values, and writes the result to the destination GPR (\code{Rd}) .
\end{adjustwidth}

\textbf{Operation}:
\begin{OpBehavior}
integer sum = UInt(GPR_val(Rs), 32) + UInt(GPR_val(Rt), 32)
GPR(Rd) = ToUBitStr(sum, 32)
PC = PC + 4
# NOTE, with 2's complement binary, it is the same for signed addition.

\end{OpBehavior}

\subsection{ AND -- And }\label{sec:insn-and}
\begin{table}[H]
\centering
\begin{isatable}
\multicolumn{ 1}{|@{}c@{} }{\small          0} & 
\multicolumn{ 1}{|@{}c@{} }{\small          0} & 
\multicolumn{ 1}{ @{}c@{} }{\small          1} &
\multicolumn{ 1}{ @{}c@{} }{\small          1} &
\multicolumn{ 1}{ @{}c@{} }{\small          0} &
\multicolumn{ 1}{ @{}c@{} }{\small          1} &
\multicolumn{ 1}{ @{}c@{} }{\small          0} &
\multicolumn{ 5}{|@{}c@{} }{\small         Rd} & 
\multicolumn{ 5}{|@{}c@{} }{\small         Rs} & 
\multicolumn{ 5}{|@{}c@{} }{\small         Rt} & 
\multicolumn{10}{|@{}c@{}|}{\small  \reserved}   
\\
\cline{1-32}
\end{isatable}
\end{table}
\vspace{-0.4cm}

\quad\textbf{Format}: \hspace{4cm} \lstinline[basicstyle=\normalsize\ttfamily]!AND Rd, Rs, Rt!

\textbf{Description:}
\begin{adjustwidth}{1cm}{1cm}
The \lstinline!AND! instruction performs a bitwise AND of two GPR (\code{Rs}, \code{Rt}) values and writes the result to the destination GPR \code{Rd}.
\end{adjustwidth}

\textbf{Operation}:
\begin{OpBehavior}
GPR(Rd) = GPR_val(Rs) & GPR_val(Rt)
PC = PC + 4

\end{OpBehavior}

\subsection{ BR -- Branch }\label{sec:insn-br}
\begin{table}[H]
\centering
\begin{isatable}
\multicolumn{ 1}{|@{}c@{} }{\small          0} & 
\multicolumn{ 1}{|@{}c@{} }{\small          0} & 
\multicolumn{ 1}{ @{}c@{} }{\small          0} &
\multicolumn{ 1}{ @{}c@{} }{\small          0} &
\multicolumn{ 1}{ @{}c@{} }{\small          0} &
\multicolumn{ 1}{ @{}c@{} }{\small          0} &
\multicolumn{ 1}{ @{}c@{} }{\small          1} &
\multicolumn{21}{|@{}c@{} }{\small      imm21} & 
\multicolumn{ 4}{|@{}c@{}|}{\small comp\_flag}   
\\
\cline{1-32}
\end{isatable}
\end{table}
\vspace{-0.4cm}

\quad\textbf{Format}: \hspace{4cm} \lstinline[basicstyle=\normalsize\ttfamily]!BR <comp_flag>, <label>!

\textbf{Description:}
\begin{adjustwidth}{1cm}{1cm}
If the specified comparison flag is `1', the \lstinline!BR! instruction changes the PC by adding an immediate offset to it. Table~\ref{tab:cmp_flags} lists all allowed comparison flags and the corresponding meaning in eQASM. <label> points to the target instruction. The assembler is responsible for converting the <label> to the immediate value \code{Imm21} according to the relative position of the target instruction and this \lstinline!BR! instruction.
\end{adjustwidth}

\textbf{Operation}:
\begin{OpBehavior}
if CompFlag_val(comp_flag) == '1':
    integer signed_sum = SInt(PC, 17) + SInt(Imm21[14:0] << 2, 17)
    PC = ToSBitStr(signed_sum, 18)[16:0]
end if

\end{OpBehavior}

\subsection{ CMP -- Compare }\label{sec:insn-cmp}
\begin{table}[H]
\centering
\begin{isatable}
\multicolumn{ 1}{|@{}c@{} }{\small          0} & 
\multicolumn{ 1}{|@{}c@{} }{\small          0} & 
\multicolumn{ 1}{ @{}c@{} }{\small          0} &
\multicolumn{ 1}{ @{}c@{} }{\small          1} &
\multicolumn{ 1}{ @{}c@{} }{\small          1} &
\multicolumn{ 1}{ @{}c@{} }{\small          0} &
\multicolumn{ 1}{ @{}c@{} }{\small          1} &
\multicolumn{ 5}{|@{}c@{} }{\small  \reserved} & 
\multicolumn{ 5}{|@{}c@{} }{\small         Rs} & 
\multicolumn{ 5}{|@{}c@{} }{\small         Rt} & 
\multicolumn{10}{|@{}c@{}|}{\small  \reserved}   
\\
\cline{1-32}
\end{isatable}
\end{table}
\vspace{-0.4cm}

\quad\textbf{Format}: \hspace{4cm} \lstinline[basicstyle=\normalsize\ttfamily]!CMP Rs, Rt!

\textbf{Description:}
\begin{adjustwidth}{1cm}{1cm}
The \lstinline!CMP! instruction compares the value of two GPRs (\code{Rs}, \code{Rt}), and updates the comparison flags based on the results.
\end{adjustwidth}

\textbf{Operation}:
\begin{OpBehavior}
COMPFLAG.ALWAYS = 1
COMPFLAG.NEVER = 0
COMPFLAG.EQ  = (GPR_val(Rt) == GPR_val(Rs))
COMPFLAG.NE  = (GPR_val(Rt) != GPR_val(Rs))
COMPFLAG.LTU = (UInt(GPR_val(Rt), 32) < UInt(GPR_val(Rs), 32))
COMPFLAG.GEU = (UInt(GPR_val(Rt), 32) >= UInt(GPR_val(Rs), 32))
COMPFLAG.LEU = (UInt(GPR_val(Rt), 32) <= UInt(GPR_val(Rs), 32))
COMPFLAG.GTU = (UInt(GPR_val(Rt), 32) > UInt(GPR_val(Rs), 32))
COMPFLAG.LT  = (SInt(GPR_val(Rt), 32) < SInt(GPR_val(Rs), 32))
COMPFLAG.GE  = (SInt(GPR_val(Rt), 32) >= SInt(GPR_val(Rs), 32))
COMPFLAG.LE  = (SInt(GPR_val(Rt), 32) <= SInt(GPR_val(Rs), 32))
COMPFLAG.GT  = (SInt(GPR_val(Rt), 32) > SInt(GPR_val(Rs), 32))
PC = PC + 4

\end{OpBehavior}

\subsection{ FBR -- Fetch Branch Register (Comparison Flag) }\label{sec:insn-fbr}
\begin{table}[H]
\centering
\begin{isatable}
\multicolumn{ 1}{|@{}c@{} }{\small          0} & 
\multicolumn{ 1}{|@{}c@{} }{\small          0} & 
\multicolumn{ 1}{ @{}c@{} }{\small          1} &
\multicolumn{ 1}{ @{}c@{} }{\small          0} &
\multicolumn{ 1}{ @{}c@{} }{\small          1} &
\multicolumn{ 1}{ @{}c@{} }{\small          0} &
\multicolumn{ 1}{ @{}c@{} }{\small          0} &
\multicolumn{ 5}{|@{}c@{} }{\small         Rd} & 
\multicolumn{16}{|@{}c@{} }{\small  \reserved} & 
\multicolumn{ 4}{|@{}c@{}|}{\small comp\_flag}   
\\
\cline{1-32}
\end{isatable}
\end{table}
\vspace{-0.4cm}

\quad\textbf{Format}: \hspace{4cm} \lstinline[basicstyle=\normalsize\ttfamily]!FBR <comp_flag>, Rd!

\textbf{Description:}
\begin{adjustwidth}{1cm}{1cm}
The \lstinline!FBR! instruction fetches the value of the given comparison flag \code{comp\_flag} and writes it to the destination GPR \code{Rd}. Table~\ref{tab:cmp_flags} lists all allowed comparison flags and the corresponding meaning in eQASM.
\end{adjustwidth}

\textbf{Operation}:
\begin{OpBehavior}
GPR(Rd) = ZeroExt(CompFlag_val(comp_flag), 32)
PC = PC + 4

\end{OpBehavior}

\subsection{ FMR -- Fetch Measurement Result }\label{sec:insn-fmr}
\begin{table}[H]
\centering
\begin{isatable}
\multicolumn{ 1}{|@{}c@{} }{\small          0} & 
\multicolumn{ 1}{|@{}c@{} }{\small          0} & 
\multicolumn{ 1}{ @{}c@{} }{\small          1} &
\multicolumn{ 1}{ @{}c@{} }{\small          0} &
\multicolumn{ 1}{ @{}c@{} }{\small          1} &
\multicolumn{ 1}{ @{}c@{} }{\small          0} &
\multicolumn{ 1}{ @{}c@{} }{\small          1} &
\multicolumn{ 5}{|@{}c@{} }{\small         Rd} & 
\multicolumn{17}{|@{}c@{} }{\small  \reserved} & 
\multicolumn{ 3}{|@{}c@{}|}{\small         Qi}   
\\
\cline{1-32}
\end{isatable}
\end{table}
\vspace{-0.4cm}

\quad\textbf{Format}: \hspace{4cm} \lstinline[basicstyle=\normalsize\ttfamily]!FMR Rd, Qi!

\textbf{Description:}
\begin{adjustwidth}{1cm}{1cm}
The \lstinline!FMR! instruction fetches the measurement result of the \textbf{last measurement instruction on qubit \code{i}} and writes it to the destination GPR \code{Rd}.
\end{adjustwidth}

\textbf{Operation}:
\begin{OpBehavior}
Wait until the last measurement instruction on qubit \code{i} finishes, i.e., the qubit measurement result register \code{Qi} gets valid, then perform the following:
GPR(Rd) = ToUBitStr(Qi, 32)
PC = PC + 4

\end{OpBehavior}

\subsection{ LD -- Load Word from Memory }\label{sec:insn-ld}
\begin{table}[H]
\centering
\begin{isatable}
\multicolumn{ 1}{|@{}c@{} }{\small          0} & 
\multicolumn{ 1}{|@{}c@{} }{\small          0} & 
\multicolumn{ 1}{ @{}c@{} }{\small          0} &
\multicolumn{ 1}{ @{}c@{} }{\small          1} &
\multicolumn{ 1}{ @{}c@{} }{\small          0} &
\multicolumn{ 1}{ @{}c@{} }{\small          0} &
\multicolumn{ 1}{ @{}c@{} }{\small          1} &
\multicolumn{ 5}{|@{}c@{} }{\small         Rd} & 
\multicolumn{ 5}{|@{}c@{} }{\small  \reserved} & 
\multicolumn{ 5}{|@{}c@{} }{\small         Rt} & 
\multicolumn{10}{|@{}c@{}|}{\small      imm10}   
\\
\cline{1-32}
\end{isatable}
\end{table}
\vspace{-0.4cm}

\quad\textbf{Format}: \hspace{4cm} \lstinline[basicstyle=\normalsize\ttfamily]!LD Rd, Rt(Imm10)!

\textbf{Description:}
\begin{adjustwidth}{1cm}{1cm}
The \lstinline!LD! instruction loads the word from the memory address specified by the register \code{Rt} with an offset \code{Imm10} into the destination GPR \code{Rd}.
\end{adjustwidth}

\textbf{Operation}:
\begin{OpBehavior}
GPR(Rd) = MemWord_val(UInt(GPR_val(Rt), 32) + SignExt(Imm10, 32))
PC = PC + 4

\end{OpBehavior}

\subsection{ LDI -- Load Immediate }\label{sec:insn-ldi}
\begin{table}[H]
\centering
\begin{isatable}
\multicolumn{ 1}{|@{}c@{} }{\small          0} & 
\multicolumn{ 1}{|@{}c@{} }{\small          0} & 
\multicolumn{ 1}{ @{}c@{} }{\small          1} &
\multicolumn{ 1}{ @{}c@{} }{\small          0} &
\multicolumn{ 1}{ @{}c@{} }{\small          1} &
\multicolumn{ 1}{ @{}c@{} }{\small          1} &
\multicolumn{ 1}{ @{}c@{} }{\small          0} &
\multicolumn{ 5}{|@{}c@{} }{\small         Rd} & 
\multicolumn{20}{|@{}c@{}|}{\small      imm20}   
\\
\cline{1-32}
\end{isatable}
\end{table}
\vspace{-0.4cm}

\quad\textbf{Format}: \hspace{4cm} \lstinline[basicstyle=\normalsize\ttfamily]!LDI Rd, Imm20!

\textbf{Description:}
\begin{adjustwidth}{1cm}{1cm}
The \lstinline!LDI! instruction loads the signed immediate value \code{Imm20} into the destination GPR \code{Rd}.
\end{adjustwidth}

\textbf{Operation}:
\begin{OpBehavior}
GPR(Rd) = SignExt(Imm20, 32)
PC = PC + 4

\end{OpBehavior}

\subsection{ LDUI -- Load Unsigned Immediate }\label{sec:insn-ldui}
\begin{table}[H]
\centering
\begin{isatable}
\multicolumn{ 1}{|@{}c@{} }{\small          0} & 
\multicolumn{ 1}{|@{}c@{} }{\small          0} & 
\multicolumn{ 1}{ @{}c@{} }{\small          1} &
\multicolumn{ 1}{ @{}c@{} }{\small          0} &
\multicolumn{ 1}{ @{}c@{} }{\small          1} &
\multicolumn{ 1}{ @{}c@{} }{\small          1} &
\multicolumn{ 1}{ @{}c@{} }{\small          1} &
\multicolumn{ 5}{|@{}c@{} }{\small         Rd} & 
\multicolumn{ 5}{|@{}c@{} }{\small         Rs} & 
\multicolumn{15}{|@{}c@{}|}{\small      imm15}   
\\
\cline{1-32}
\end{isatable}
\end{table}
\vspace{-0.4cm}

\quad\textbf{Format}: \hspace{4cm} \lstinline[basicstyle=\normalsize\ttfamily]!LDUI Rd, Rs, Imm15!

\textbf{Description:}
\begin{adjustwidth}{1cm}{1cm}
The \lstinline!LDUI! instruction inserts an 15-bit constant into the upper 15 bits of the destination GPR \code{Rd}.
\end{adjustwidth}

\textbf{Operation}:
\begin{OpBehavior}
GPR(Rd) = Imm15 << 17 | GPR_val(Rs)[16:0]
PC = PC + 4

\end{OpBehavior}

\subsection{ NOP -- No Operation }\label{sec:insn-nop}
\begin{table}[H]
\centering
\begin{isatable}
\multicolumn{ 1}{|@{}c@{} }{\small          0} & 
\multicolumn{ 1}{|@{}c@{} }{\small          0} & 
\multicolumn{ 1}{ @{}c@{} }{\small          0} &
\multicolumn{ 1}{ @{}c@{} }{\small          0} &
\multicolumn{ 1}{ @{}c@{} }{\small          0} &
\multicolumn{ 1}{ @{}c@{} }{\small          0} &
\multicolumn{ 1}{ @{}c@{} }{\small          0} &
\multicolumn{ 1}{|@{}c@{} }{\small          0} & 
\multicolumn{ 1}{ @{}c@{} }{\small          0} &
\multicolumn{ 1}{ @{}c@{} }{\small          0} &
\multicolumn{ 1}{ @{}c@{} }{\small          0} &
\multicolumn{ 1}{ @{}c@{} }{\small          0} &
\multicolumn{ 1}{ @{}c@{} }{\small          0} &
\multicolumn{ 1}{ @{}c@{} }{\small          0} &
\multicolumn{ 1}{ @{}c@{} }{\small          0} &
\multicolumn{ 1}{ @{}c@{} }{\small          0} &
\multicolumn{ 1}{ @{}c@{} }{\small          0} &
\multicolumn{ 1}{ @{}c@{} }{\small          0} &
\multicolumn{ 1}{ @{}c@{} }{\small          0} &
\multicolumn{ 1}{ @{}c@{} }{\small          0} &
\multicolumn{ 1}{ @{}c@{} }{\small          0} &
\multicolumn{ 1}{ @{}c@{} }{\small          0} &
\multicolumn{ 1}{ @{}c@{} }{\small          0} &
\multicolumn{ 1}{ @{}c@{} }{\small          0} &
\multicolumn{ 1}{ @{}c@{} }{\small          0} &
\multicolumn{ 1}{ @{}c@{} }{\small          0} &
\multicolumn{ 1}{ @{}c@{} }{\small          0} &
\multicolumn{ 1}{ @{}c@{} }{\small          0} &
\multicolumn{ 1}{|@{}c@{} }{\small          0} & 
\multicolumn{ 1}{ @{}c@{} }{\small          0} &
\multicolumn{ 1}{ @{}c@{} }{\small          0} &
\multicolumn{ 1}{ @{}c@{}|}{\small          0}
\\
\cline{1-32}
\end{isatable}
\end{table}
\vspace{-0.4cm}

\quad\textbf{Format}: \hspace{4cm} \lstinline[basicstyle=\normalsize\ttfamily]!NOP!

\textbf{Description:}
\begin{adjustwidth}{1cm}{1cm}
The \lstinline!NOP! instruction performs no operation.
\end{adjustwidth}

\textbf{Operation}:
\begin{OpBehavior}
PC = PC + 4

\end{OpBehavior}

\subsection{ NOT -- Not }\label{sec:insn-not}
\begin{table}[H]
\centering
\begin{isatable}
\multicolumn{ 1}{|@{}c@{} }{\small          0} & 
\multicolumn{ 1}{|@{}c@{} }{\small          0} & 
\multicolumn{ 1}{ @{}c@{} }{\small          1} &
\multicolumn{ 1}{ @{}c@{} }{\small          1} &
\multicolumn{ 1}{ @{}c@{} }{\small          0} &
\multicolumn{ 1}{ @{}c@{} }{\small          1} &
\multicolumn{ 1}{ @{}c@{} }{\small          1} &
\multicolumn{ 5}{|@{}c@{} }{\small         Rd} & 
\multicolumn{ 5}{|@{}c@{} }{\small  \reserved} & 
\multicolumn{ 5}{|@{}c@{} }{\small         Rt} & 
\multicolumn{10}{|@{}c@{}|}{\small  \reserved}   
\\
\cline{1-32}
\end{isatable}
\end{table}
\vspace{-0.4cm}

\quad\textbf{Format}: \hspace{4cm} \lstinline[basicstyle=\normalsize\ttfamily]!NOT Rd, Rt!

\textbf{Description:}
\begin{adjustwidth}{1cm}{1cm}
The \lstinline!NOT! instruction performs a bitwise NOT of a GPR (\code{Rt}) value and writes the result to the destination GPR \code{Rd}.
\end{adjustwidth}

\textbf{Operation}:
\begin{OpBehavior}
GPR(Rd) = ~GPR_val(Rt)
PC = PC + 4

\end{OpBehavior}

\subsection{ OR -- Or }\label{sec:insn-or}
\begin{table}[H]
\centering
\begin{isatable}
\multicolumn{ 1}{|@{}c@{} }{\small          0} & 
\multicolumn{ 1}{|@{}c@{} }{\small          0} & 
\multicolumn{ 1}{ @{}c@{} }{\small          1} &
\multicolumn{ 1}{ @{}c@{} }{\small          1} &
\multicolumn{ 1}{ @{}c@{} }{\small          0} &
\multicolumn{ 1}{ @{}c@{} }{\small          0} &
\multicolumn{ 1}{ @{}c@{} }{\small          0} &
\multicolumn{ 5}{|@{}c@{} }{\small         Rd} & 
\multicolumn{ 5}{|@{}c@{} }{\small         Rs} & 
\multicolumn{ 5}{|@{}c@{} }{\small         Rt} & 
\multicolumn{10}{|@{}c@{}|}{\small  \reserved}   
\\
\cline{1-32}
\end{isatable}
\end{table}
\vspace{-0.4cm}

\quad\textbf{Format}: \hspace{4cm} \lstinline[basicstyle=\normalsize\ttfamily]!OR Rd, Rs, Rt!

\textbf{Description:}
\begin{adjustwidth}{1cm}{1cm}
The \lstinline!OR! instruction performs a bitwise OR of two GPR (\code{Rs}, \code{Rt}) values and writes the result to the destination GPR \code{Rd}.
\end{adjustwidth}

\textbf{Operation}:
\begin{OpBehavior}
GPR(Rd) = GPR_val(Rs) | GPR_val(Rt)
PC = PC + 4

\end{OpBehavior}

\subsection{ QWAIT -- Quantum Wait Immediate }\label{sec:insn-qwait}
\begin{table}[H]
\centering
\begin{isatable}
\multicolumn{ 1}{|@{}c@{} }{\small          0} & 
\multicolumn{ 1}{|@{}c@{} }{\small          1} & 
\multicolumn{ 1}{ @{}c@{} }{\small          0} &
\multicolumn{ 1}{ @{}c@{} }{\small          0} &
\multicolumn{ 1}{ @{}c@{} }{\small          0} &
\multicolumn{ 1}{ @{}c@{} }{\small          0} &
\multicolumn{ 1}{ @{}c@{} }{\small          0} &
\multicolumn{ 5}{|@{}c@{} }{\small  \reserved} & 
\multicolumn{20}{|@{}c@{}|}{\small      Imm20}   
\\
\cline{1-32}
\end{isatable}
\end{table}
\vspace{-0.4cm}

\quad\textbf{Format}: \hspace{4cm} \lstinline[basicstyle=\normalsize\ttfamily]!QWAIT Imm20!

\textbf{Description:}
\begin{adjustwidth}{1cm}{1cm}
The \lstinline!QWAIT! instruction creates a new timing point with a new timing label which is \code{Imm20} cycles later than the previous timing point.
\end{adjustwidth}

\textbf{Operation}:
\begin{OpBehavior}
TimingLabel = TimingLabel + 1
TimingQueue.push(TimingLabel, Imm20)
PC = PC + 4

\end{OpBehavior}

\subsection{ QWAITR -- Quantum Wait Register }\label{sec:insn-qwaitr}
\begin{table}[H]
\centering
\begin{isatable}
\multicolumn{ 1}{|@{}c@{} }{\small          0} & 
\multicolumn{ 1}{|@{}c@{} }{\small          1} & 
\multicolumn{ 1}{ @{}c@{} }{\small          0} &
\multicolumn{ 1}{ @{}c@{} }{\small          0} &
\multicolumn{ 1}{ @{}c@{} }{\small          0} &
\multicolumn{ 1}{ @{}c@{} }{\small          0} &
\multicolumn{ 1}{ @{}c@{} }{\small          1} &
\multicolumn{ 5}{|@{}c@{} }{\small  \reserved} & 
\multicolumn{ 5}{|@{}c@{} }{\small         Rs} & 
\multicolumn{15}{|@{}c@{}|}{\small  \reserved}   
\\
\cline{1-32}
\end{isatable}
\end{table}
\vspace{-0.4cm}

\quad\textbf{Format}: \hspace{4cm} \lstinline[basicstyle=\normalsize\ttfamily]!QWAITR Rs!

\textbf{Description:}
\begin{adjustwidth}{1cm}{1cm}
The \lstinline!QWAITR! instruction creates a new timing point with a new timing label which is $k$ cycles later than the previous timing point. $k$ is an unsigned value specified by the 20 least significant bits of GPR \code{Rs}.
\end{adjustwidth}

\textbf{Operation}:
\begin{OpBehavior}
TimingLabel = TimingLabel + 1
TimingQueue.push(TimingLabel, GPR_val(Rs)[19:0])
PC = PC + 4

\end{OpBehavior}

\input{isa/smi.tex}

\subsection{ ST -- Store Word to Memory }\label{sec:insn-st}
\begin{table}[H]
\centering
\begin{isatable}
\multicolumn{ 1}{|@{}c@{} }{\small          0} & 
\multicolumn{ 1}{|@{}c@{} }{\small          0} & 
\multicolumn{ 1}{ @{}c@{} }{\small          0} &
\multicolumn{ 1}{ @{}c@{} }{\small          1} &
\multicolumn{ 1}{ @{}c@{} }{\small          0} &
\multicolumn{ 1}{ @{}c@{} }{\small          1} &
\multicolumn{ 1}{ @{}c@{} }{\small          0} &
\multicolumn{ 5}{|@{}c@{} }{\small  \reserved} & 
\multicolumn{ 5}{|@{}c@{} }{\small         Rs} & 
\multicolumn{ 5}{|@{}c@{} }{\small         Rt} & 
\multicolumn{10}{|@{}c@{}|}{\small      imm10}   
\\
\cline{1-32}
\end{isatable}
\end{table}
\vspace{-0.4cm}

\quad\textbf{Format}: \hspace{4cm} \lstinline[basicstyle=\normalsize\ttfamily]!ST Rs, Rt(Imm10)!

\textbf{Description:}
\begin{adjustwidth}{1cm}{1cm}
The \lstinline!ST! instruction stores the value of GPR \code{Rs} to the memory address specified by the register \code{Rt} with an offset \code{Imm10}.
\end{adjustwidth}

\textbf{Operation}:
\begin{OpBehavior}
MemWord(UInt(GPR_val(Rt), 32) + SignExt(Imm10, 32)) = GPR_val(Rs)
PC = PC + 4

\end{OpBehavior}

\subsection{ STOP -- Stop }\label{sec:insn-stop}
\begin{table}[H]
\centering
\begin{isatable}
\multicolumn{ 1}{|@{}c@{} }{\small          0} & 
\multicolumn{ 1}{|@{}c@{} }{\small          0} & 
\multicolumn{ 1}{ @{}c@{} }{\small          0} &
\multicolumn{ 1}{ @{}c@{} }{\small          1} &
\multicolumn{ 1}{ @{}c@{} }{\small          0} &
\multicolumn{ 1}{ @{}c@{} }{\small          0} &
\multicolumn{ 1}{ @{}c@{} }{\small          0} &
\multicolumn{25}{|@{}c@{}|}{\small  \reserved}   
\\
\cline{1-32}
\end{isatable}
\end{table}
\vspace{-0.4cm}

\quad\textbf{Format}: \hspace{4cm} \lstinline[basicstyle=\normalsize\ttfamily]!STOP!

\textbf{Description:}
\begin{adjustwidth}{1cm}{1cm}
The \lstinline!STOP! instruction sets the execution flag \code{STOP}, and repeats executing itself infinitely. In other words, it stops the processor.
\end{adjustwidth}

\textbf{Operation}:
\begin{OpBehavior}
EXEFLAG.STOP = 1
PC = PC

\end{OpBehavior}

\subsection{ SUB -- Subtraction }\label{sec:insn-sub}
\begin{table}[H]
\centering
\begin{isatable}
\multicolumn{ 1}{|@{}c@{} }{\small          0} & 
\multicolumn{ 1}{|@{}c@{} }{\small          0} & 
\multicolumn{ 1}{ @{}c@{} }{\small          1} &
\multicolumn{ 1}{ @{}c@{} }{\small          1} &
\multicolumn{ 1}{ @{}c@{} }{\small          1} &
\multicolumn{ 1}{ @{}c@{} }{\small          1} &
\multicolumn{ 1}{ @{}c@{} }{\small          1} &
\multicolumn{ 5}{|@{}c@{} }{\small         Rd} & 
\multicolumn{ 5}{|@{}c@{} }{\small         Rs} & 
\multicolumn{ 5}{|@{}c@{} }{\small         Rt} & 
\multicolumn{10}{|@{}c@{}|}{\small  \reserved}   
\\
\cline{1-32}
\end{isatable}
\end{table}
\vspace{-0.4cm}

\quad\textbf{Format}: \hspace{4cm} \lstinline[basicstyle=\normalsize\ttfamily]!SUB Rd, Rs, Rt!

\textbf{Description:}
\begin{adjustwidth}{1cm}{1cm}
The \lstinline!SUB! instruction subtract a GPR (\code{Rs}) value from another GPR (\code{Rt}) value, and writes the result to the destination GPR (\code{Rd}).
\end{adjustwidth}

\textbf{Operation}:
\begin{OpBehavior}
integer sum = UInt(GPR_val(Rs), 32) + UInt(NOT(GPR_val(Rt)), 32) + UInt(1, 32)
GPR(Rd) = ToUBitStr(sum, 32)
PC = PC + 4
# NOTE, with 2's complement binary, it is the same for signed subtraction.

\end{OpBehavior}

\subsection{ XOR -- Exclusive Or }\label{sec:insn-xor}
\begin{table}[H]
\centering
\begin{isatable}
\multicolumn{ 1}{|@{}c@{} }{\small          0} & 
\multicolumn{ 1}{|@{}c@{} }{\small          0} & 
\multicolumn{ 1}{ @{}c@{} }{\small          1} &
\multicolumn{ 1}{ @{}c@{} }{\small          1} &
\multicolumn{ 1}{ @{}c@{} }{\small          0} &
\multicolumn{ 1}{ @{}c@{} }{\small          0} &
\multicolumn{ 1}{ @{}c@{} }{\small          1} &
\multicolumn{ 5}{|@{}c@{} }{\small         Rd} & 
\multicolumn{ 5}{|@{}c@{} }{\small         Rs} & 
\multicolumn{ 5}{|@{}c@{} }{\small         Rt} & 
\multicolumn{10}{|@{}c@{}|}{\small  \reserved}   
\\
\cline{1-32}
\end{isatable}
\end{table}
\vspace{-0.4cm}

\quad\textbf{Format}: \hspace{4cm} \lstinline[basicstyle=\normalsize\ttfamily]!XOR Rd, Rs, Rt!

\textbf{Description:}
\begin{adjustwidth}{1cm}{1cm}
The \lstinline!XOR! instruction performs a bitwise XOR of two GPR (\code{Rs}, \code{Rt}) values and writes the result to the destination GPR \code{Rd}.
\end{adjustwidth}

\textbf{Operation}:
\begin{OpBehavior}
GPR(Rd) = GPR_val(Rs) ^ GPR_val(Rt)
PC = PC + 4
\end{OpBehavior}

%% file: isa/smi.tex
\subsection{ SMIS -- Set Mask Immediate for Singe-qubit Operations }\label{sec:insn-smis}
\begin{table}[H]
\centering
\begin{isatable}
\multicolumn{ 1}{|@{}c@{} }{\small          0} & 
\multicolumn{ 1}{|@{}c@{} }{\small          1} & 
\multicolumn{ 1}{ @{}c@{} }{\small          0} &
\multicolumn{ 1}{ @{}c@{} }{\small          0} &
\multicolumn{ 1}{ @{}c@{} }{\small          0} &
\multicolumn{ 1}{ @{}c@{} }{\small          0} &
\multicolumn{ 1}{ @{}c@{} }{\small          0} &
\multicolumn{ 6}{|@{}c@{} }{\small         Sd} & 
\multicolumn{12}{|@{}c@{} }{\small  \reserved} & 
\multicolumn{ 7}{|@{}c@{}|}{\small       Imm7}   
\\
\cline{1-32}
\end{isatable}
\end{table}
\vspace{-0.4cm}

\quad\textbf{Format}: \hspace{4cm} \lstinline[basicstyle=\normalsize\ttfamily]!SMIS Sd, <Qubit List>!
\begin{adjustwidth}{1cm}{1cm}
where, \code{<Qubit List>}  has the following format:
\begin{center}
    \lstinline[basicstyle=\normalsize\ttfamily]!{<qubit address>[, <qubit address>]*}!
\end{center}
\end{adjustwidth}

\begin{figure}[bt]
    \centering
    \hspace{-1cm}
    \includegraphics[width=0.6\columnwidth]{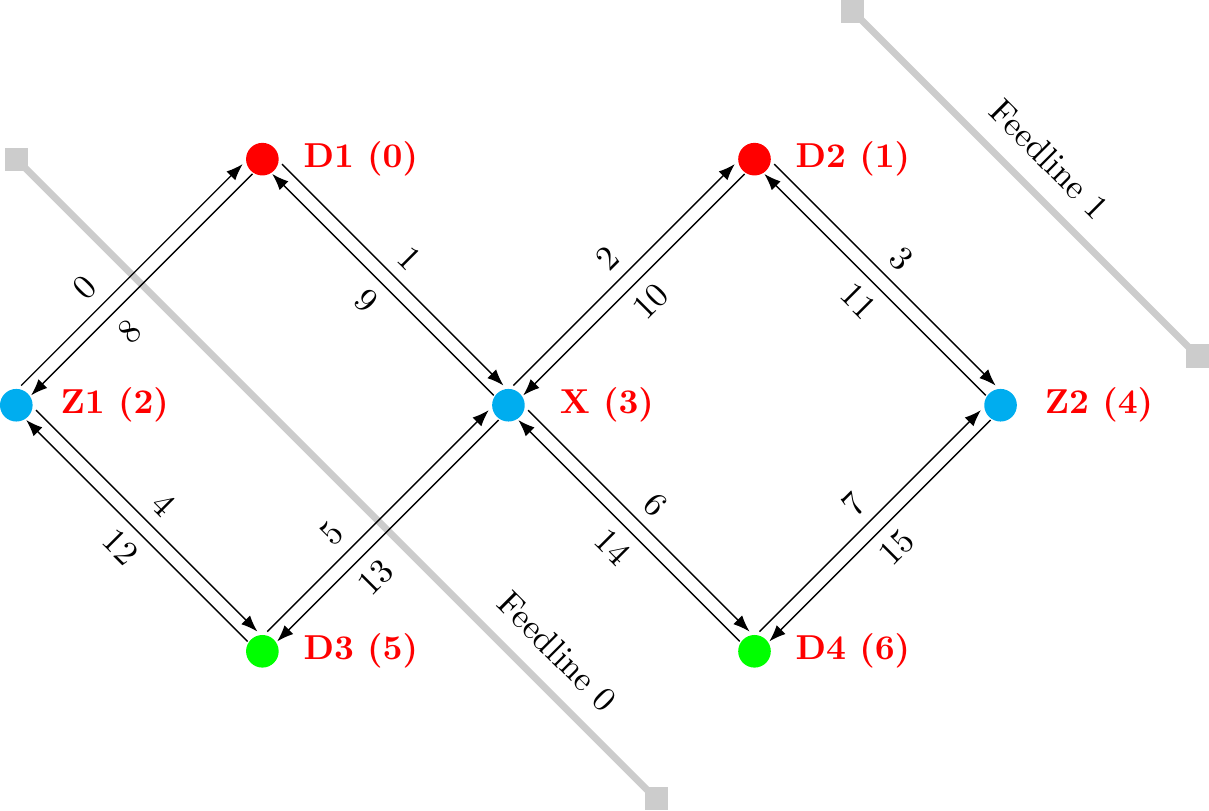}
    \caption{The ordering of individual qubits and allowed qubit pairs.}
    \label{fig:order}
\end{figure}

\textbf{Description:}
\begin{adjustwidth}{1cm}{1cm}
In CC-Light eQASM, the single-qubit target registers use a mask format in the binary. Each bit in the mask of the value \bin{1} indicates that the corresponding qubit is selected (see Fig.~\ref{fig:order}).
The \lstinline!SMIS! instruction sets the single-qubit quantum operation target register \code{Sd} to the mask which selects all qubits as listed in the set \code{<Qubit List>}. The assembler is responsible for translating \code{<Qubit List}> into the mask \code{Imm7}.

\end{adjustwidth}

\textbf{Operation}:
\begin{OpBehavior}
QOTRS(Sd) = Imm7
PC = PC + 4

\end{OpBehavior}

\subsection{ SMIT -- Set Mask Immediate for Two-qubit Operations }\label{sec:insn-smit}
\begin{table}[H]
\centering
\begin{isatable}
\multicolumn{ 1}{|@{}c@{} }{\small          0} & 
\multicolumn{ 1}{|@{}c@{} }{\small          1} & 
\multicolumn{ 1}{ @{}c@{} }{\small          0} &
\multicolumn{ 1}{ @{}c@{} }{\small          1} &
\multicolumn{ 1}{ @{}c@{} }{\small          0} &
\multicolumn{ 1}{ @{}c@{} }{\small          0} &
\multicolumn{ 1}{ @{}c@{} }{\small          0} &
\multicolumn{ 6}{|@{}c@{} }{\small         Td} & 
\multicolumn{ 3}{|@{}c@{} }{\small  \reserved} & 
\multicolumn{16}{|@{}c@{}|}{\small      Imm16}   
\\
\cline{1-32}
\end{isatable}
\end{table}
\vspace{-0.4cm}

\quad\textbf{Format}: \hspace{4cm} \lstinline[basicstyle=\normalsize\ttfamily]!SMIT Td, <Qubit Pair List>!
\begin{adjustwidth}{1cm}{1cm}
where,  \code{<Qubit Pair List>}  has the following format:
\begin{center}
\lstinline[basicstyle=\normalsize\ttfamily]!{<Qubit Pair>[, <Qubit Pair>]*}!
\end{center}
where, \code{<Qubit Pair>}  has the following format:
\begin{center}
\lstinline[basicstyle=\normalsize\ttfamily]!(<Source Qubit Address>, <Target Qubit Address>)!
\end{center}
\end{adjustwidth}
\textbf{Description:}
\begin{adjustwidth}{1cm}{1cm}
In CC-Light eQASM, the two-qubit target registers use a mask format in the binary. Each bit in the mask of the value \bin{1} indicates one of 16 allowed qubit pairs selected (see Fig.~\ref{fig:order}).
The \lstinline!SMIT! instruction sets the two-qubit quantum operation target register \code{Td} to the mask which selects all allowed qubit pairs as listed in the set \code{<Qubit Pair List>}. The assembler is responsible for translating \code{<Qubit Pair List}> into the mask \code{Imm16}.

\end{adjustwidth}

\textbf{Operation}:
\begin{OpBehavior}
QOTRT(Td) = Imm16
PC = PC + 4
\end{OpBehavior}

%% file: app-01-examples.tex
\section{Examples}
In this section, we give some examples to show how to use eQASM to describe some quantum experiments and quantum algorithms.

\subsection{Quantum Experiments}
eQASM targets nowadays and near-term devices. Since calibration occupies most of the time when the quantum chip is being used, eQASM should also support the quantum experiments to calibrate qubits. This subsection shows the eQASM program of some widely-used quantum experiments, including $T_1$, Rabi, \textit{AllXY}, etc.

\subsubsection{$T_1$} In this experiment, the CTPG only requires the \qgate{X} pulse being uploaded.

\begin{lstlisting}
.def_sym init_waiting_time  10000
.def_sym msmt_duration      15

.register r0 num_repitition
.register r1 max_repitition
LDI max_repitition,         10000

.register r3 start_interval
LDI start_interval, 50      # 1 us

.register r2 sweep_step
LDI sweep_step,     50      # 1 us

.register r4 max_interval
LDI max_interval,   5000    # 100 us

.register r5 round_interval

.register r31 constant_one
LDI constant_one, 1

SMIS S0, {0}
LDI num_repitition, 0
Round_Start:
    LDI round_interval, start_interval

iteration_start:
    QWAIT init_waiting_time
    X S0
    QWAITR round_interval
    MEASZ S0

    # go to next iteration if not reaching the maximal interval
    add round_interval, round_interval, sweep_step
    CMP round_interval, max_interval
    NOP
    BR LTU, iteration_start

    add num_repitition, num_repitition, constant_one

    # go to next round if the repition is insufficient
    CMP num_repitition, max_repitition
    NOP
    BR LTU, round_start

\end{lstlisting}

\subsubsection{Rabi} In this experiment, the CTPG should be uploaded with multiple variants of the \qgate{X} gate, with each variant of a particular amplitude.

\lstset{matchrangestart=t}
\begin{lstlisting}[linerange={1-25, 215-225}, showlines=true]
.def_sym init_waiting_time  10000
.def_sym msmt_duration      15

.register r0 num_repitition
.register r1 max_repitition
LDI max_repitition,         1000

.register r31 constant_one
LDI constant_one, 1

SMIS S0, {0}
LDI num_repitition, 0

Round_Start:
    QWAIT       init_waiting_time
    X_Amp_0     S0
    QWAITR      round_interval
    MEASZ       S0

    QWAIT       init_waiting_time
    X_Amp_1     S0
    QWAITR      round_interval
    MEASZ       S0
    # not showing the following 37 iterations here

    QWAIT       init_waiting_time
    X_Amp_39    S0
    QWAITR      round_interval
    MEASZ       S0

    # go to next round if the repition is insufficient
    add num_repitition, num_repitition, constant_one
    CMP num_repitition, max_repitition
    NOP
    BR LTU, round_start
\end{lstlisting}

\subsection{Quantum Algorithms}
This subsection shows how eQASM program can be used to express some basic quantum algorithms.

\subsubsection{Grover's Search}
To explain it in a simpler way, Grover's search algorithm is to search a specific element $x$ in a randomly ordered database $S$ such that the given function $f(x)=1$.
The number of elements in $S$ is $N$. It allows that $f(x)=1$ can have multiple solutions $\{x_1, x_2, \ldots, x_n\}$ in $S$, and Grover's search will finally return a random one of them. But it requires that $n \le \floor{N/2}$ to make sure Grover's search can work.

\begin{figure}[H]
    \centering
\begin{equation*}
\Qcircuit @C=1.5em @R=1em {
\lstick{x_1 \ket{0}} & \qw &      \gate{H} & \multigate{2}{\mathcal{O}} & \gate{H} & \ctrlo{1}  & \gate{H} & \meter & \cw & \rstick{m_2}  \\
\lstick{x_0 \ket{0}} & \qw &      \gate{H} & \ghost{\mathcal{O}}        & \gate{H} & \ctrlo{-1} & \gate{H} & \meter & \cw & \rstick{m_1} \\
\lstick{y \ket{0}} & \gate{X} & \gate{H} & \ghost{\mathcal{O}}        & \qw      & \qw        & \qw      & \qw & \qw  &  \qw \gategroup{1}{5}{2}{7}{0.5em}{--}\gategroup{1}{4}{3}{7}{.8em}{\{}\gategroup{1}{4}{3}{7}{1.3em}{\}}
}
\end{equation*}
    \caption{Quantum circuit of Grover's search.}
    \label{fig:grover}
\end{figure}
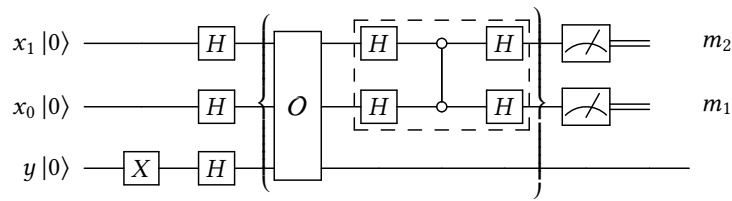

Take the quantum circuit shown in Figure~\ref{fig:grover} as an example, which implements a Grover's search over a database with $N=4$. In this case, $x$ can be represented using a binary format $\textrm{0b}x_1x_0$, where $x_1, x_0 \in\{0,1\}$, and each bit is encoded into a data qubit. $y$ is a single-bit value and can be encoded into a qubit, called ancilla qubit (labeled as $y$).
The Grover's search contains the three steps: initialization (left to the bracket), Grover iterations (inside the brackets), and measurement (right to the brackets).

The initialization process (preparing all qubits in the $\ket{0}$ state followed by a \qgate{X} gate and three parallel \qgate{H} gates) puts the data qubits and ancilla in the maximal superposition state:
\begin{align*}
    \ket{\psi _1} & = \frac{1}{2\sqrt{2}}(\ket{0} + \ket{1})\otimes(\ket{0} + \ket{1})\otimes(\ket{0} - \ket{1})\\
    & = \frac{1}{2\sqrt{2}}\left[\ket{000}-\ket{001}+\ket{010}-\ket{001}+\ket{100}-\ket{101}+\ket{110}-\ket{111}\right]
\end{align*}

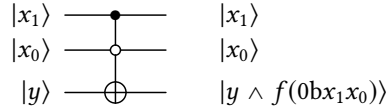
\begin{figure}[H]
    \centering
        \begin{align*}
            \Qcircuit @C=1.5em @R=1em {
                \lstick{\ket{x_1}} &  \ctrl{1}  & \qw & \rstick{\ket{x_1}}  \\
                \lstick{\ket{x_0}} &  \ctrlo{1} & \qw & \rstick{\ket{x_0}} \\
                \lstick{\ket{y}}   &  \targ     & \qw & \rstick{\ket{y \wedge f(\textrm{0b}x_1x_0)}}
            }
            \end{align*}
    \caption{Quantum circuit implementing the oracle $(\ket{x}, \ket{y})\rightarrow(\ket{x}, \ket{y \wedge f(x)})$, where $f(x)=1$ only when $x=\textrm{0b}10$.}
    \label{fig:oracle01}
\end{figure}

A Grover iteration consists of calling oracle and inversion about mean (dashed box).
The oracle $\mathcal{O}$ implements the function $(\ket{x}, \ket{y})\rightarrow(\ket{x}, \ket{y \wedge f(x)})$, where $\wedge$ is bitwise XOR on two bit strings. For example, if the given function $f(x)=1$ for $x=\textrm{0b}10$ and $f(x)=0$ for else, then the oracle  $\mathcal{O}$ can be implemented using the quantum circuit as shown in Figure~\ref{fig:oracle01}.

After the oracle, the quantum state turns into
\begin{align}
\label{eq:after_oracle}
    \ket{\psi _2}\left[(-1)^{f(0)}\ket{00} +
    (-1)^{f(1)}\ket{01} +
    (-1)^{f(2)}\ket{10} +
    (-1)^{f(3)}\ket{11}\right]\otimes(\ket{0} - \ket{1}).
\end{align}

The operator sandwiched between four \qgate{H} gates in the dashed box is an operator that only flips the phase of the $\ket{00}$ state in the superposition. This operator is one of the four physically implementable C-Phase gates $cU_{ij}$ with superconducting qubits:
\begin{align}
    cU_{ij}\ket{kl} = (-1)^{\delta_{ik}\delta_{jl}}\ket{kl}
\end{align}
where $i,j,k,l\in\{0,1\}$ and $\delta$ is Kronecker's delta.
Bearing this information in mind, we can know that the dashed box implements the inversion about mean for the input state. In other words, if the input state is
\begin{align*}
    \ket{\psi_i}=\alpha_{00}\ket{00} + \alpha_{01}\ket{01} + \alpha_{10}\ket{10} + \alpha_{00}\ket{00},
\end{align*}
then, the output state is
\begin{align*}
    \ket{\psi_o}=(2\alpha_{\mathrm{m}}- \alpha_{00}\ket{00}) + (2\alpha_{\mathrm{m}}- \alpha_{01})\ket{01} + (2\alpha_{\mathrm{m}}- \alpha_{10})\ket{10} + (2\alpha_{\mathrm{m}}- \alpha_{00})\ket{00},
\end{align*}
where $\alpha_{\mathrm{m}}=1/4(\alpha_{00}+\alpha_{01}+\alpha_{10}+\alpha_{11})$.

If $\alpha_{ij}$ is the amplitude that has an opposite sign to the other amplitudes, the data qubits state after inversion about mean turns into
\begin{align*}
    \ket{\psi _3} = \ket{ij}\otimes(\ket{0} - \ket{1}),
\end{align*}
where data qubits are in a non-superposed state ($x_1=i, x_0=j$), which can be readout to reveal the solution to the problem.

\begin{figure}[H]
    \centering
\begin{align*}
\Qcircuit @C=1.5em @R=1em {
\lstick{Q_2 \ket{0}} & \gate{H} & \multigate{1}{\mathcal{O}} & \gate{H} & \ctrlo{1}  & \gate{H} & \meter & \cw & \rstick{m_2}  \\
\lstick{Q_0 \ket{0}} & \gate{H} & \ghost{\mathcal{O}} & \gate{H} & \ctrlo{-1} & \gate{H} & \meter & \cw & \rstick{m_1}\gategroup{1}{4}{2}{6}{.5em}{--}
}
\end{align*}
    \caption{Quantum circuit of Grover's search.}
    \label{fig:2qgrover}
\end{figure}
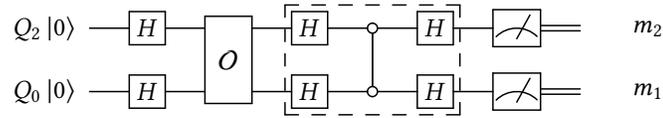

\paragraph{More Analysis}
As shown in Eq.~\ref{eq:after_oracle}, the ancilla qubit is no longer entangled with the data qubits, and the following steps only involve the first two qubits. What matters is the phase of phase of each state in the superposition. If we can prepare the state $(-1)^{f(0)}\ket{00} + (-1)^{f(1)}\ket{01} + (-1)^{f(2)}\ket{10} + (-1)^{f(3)}\ket{11}$ using a two-qubit oracle, we can simply the implementation of the algorithm from using three qubits into using two-qubits as shown in Fig.~\ref{fig:2qgrover}.

Luckily, we can use the two-qubit C-Phase gate $cU_{ij}$ to implement the oracle $\mathcal{O}$ corresponding to each case where $x=\textrm{0b}ij$ is the only solution to $f(x)=1$.

The eQASM code implementing the simplified quantum circuit of Grover's search algorithm in Fig.~\ref{fig:2qgrover} is shown in Listing~\ref{lst:grover}. Note, we replace each \qgate{H} gate with an \qgate{Y_90} gate to reduce 6 more physical gates.

\begin{lstlisting}[caption={eQASM code for the two-qubit implementation of Grover's search algorithm.}, captionpos=t, label={lst:grover}, style=eQASMStyle]
# select qubit 0, 2
# assume the given function f(x)=1 when x1,x0 = 0,1 

.def_sym init_waiting_time  10000  # 200 us
.def_sym msmt_duration      15     # 300 ns
LDI r2, 1000        # number of repetition
LDi r0, 0           # counter
LDI r1, 1

smis s7, {0, 2}
smit t0, {(0, 2)}

LoopStart:
  QWAIT   init_waiting_time   # initialize all qubits
     Y90     s7
     cU01    t0
  2, Y90     s7
     cU00    t0
  2, Y90     s7
     MeasZ   s7
  QWAIT   msmt_duration
  add r0, r0, r1
  cmp r0, r2
  BR LEU, LoopStart
\end{lstlisting}

%% file: app-02-qmap.tex
\section{.QMAP File Specification}
\label{sec:qmap}

An example \filename{qisa\_opcode.qmap} file is shown in Listing~\ref{lst:ex_qmap} which is being used in our ongoing experiments. When the assembler converts CC-Light eQASM assembly code into binary code, this file will be read by the assembler to generate the opcode of each quantum operation. Note, to enable an easy debugging, the programmer uses the operation names such as \code{cw\_01} instead of \code{X} at the assembly level in this configuration.

\begin{lstlisting}[caption={Example \filename{qisa\_opcode.qmap} accepted by the assembler.},captionpos=b, label={lst:ex_qmap}, language=Python]
# Quantum Instructions (double instruction format)

# No arguments
def_q_arg_none["qnop"]    = 0x00

# Single-qubit operations using the 'S' registers

# Initializing qubit by idling
def_q_arg_st["prepz"]	    = 0x2

# Measurements
#  reserved msmt            = 0x04
#  reserved msmt            = 0x05
def_q_arg_st['MeasZ']       = 0x06
#  reserved msmt            = 0x07

# Microwave operations require a codeword from 1 to 127 except 4~7.
def_q_arg_st["cw_00"]	    = 0x8
def_q_arg_st["cw_01"]	    = 0x9
def_q_arg_st["cw_02"]	    = 0xa
def_q_arg_st["cw_03"]	    = 0xb
def_q_arg_st["cw_04"]	    = 0xc
def_q_arg_st["cw_05"]	    = 0xd
def_q_arg_st["cw_06"]	    = 0xe
def_q_arg_st["cw_07"]	    = 0xf
def_q_arg_st["cw_08"]	    = 0x10
def_q_arg_st["cw_09"]	    = 0x11
def_q_arg_st["cw_10"]	    = 0x12
def_q_arg_st["cw_11"]	    = 0x13
def_q_arg_st["cw_12"]	    = 0x14
def_q_arg_st["cw_13"]	    = 0x15
def_q_arg_st["cw_14"]	    = 0x16
def_q_arg_st["cw_15"]	    = 0x17
def_q_arg_st["cw_16"]	    = 0x18
def_q_arg_st["cw_17"]	    = 0x19
def_q_arg_st["cw_18"]	    = 0x1a
def_q_arg_st["cw_19"]	    = 0x1b
def_q_arg_st["cw_20"]	    = 0x1c
def_q_arg_st["cw_21"]	    = 0x1d
def_q_arg_st["cw_22"]	    = 0x1e
def_q_arg_st["cw_23"]	    = 0x1f
def_q_arg_st["cw_24"]	    = 0x20
def_q_arg_st["cw_25"]	    = 0x21
def_q_arg_st["cw_26"]	    = 0x22
def_q_arg_st["cw_27"]	    = 0x23
def_q_arg_st["cw_28"]	    = 0x24
def_q_arg_st["cw_29"]	    = 0x25
def_q_arg_st["cw_30"]	    = 0x26
def_q_arg_st["cw_31"]	    = 0x27
def_q_arg_st["C0_cw_00"]	= 0x30
def_q_arg_st["C0_cw_01"]	= 0x31
def_q_arg_st["C0_cw_02"]	= 0x32
def_q_arg_st["C0_cw_03"]	= 0x33
def_q_arg_st["C0_cw_04"]	= 0x34
def_q_arg_st["C0_cw_05"]	= 0x35
def_q_arg_st["C0_cw_06"]	= 0x36
def_q_arg_st["C0_cw_07"]	= 0x37
def_q_arg_st["C0_cw_08"]	= 0x38
def_q_arg_st["C1_cw_00"]	= 0x28
def_q_arg_st["C1_cw_07"]	= 0x2f
def_q_arg_st["C1_cw_01"]	= 0x29
def_q_arg_st["C1_cw_02"]	= 0x2a
def_q_arg_st["C1_cw_03"]	= 0x2b
def_q_arg_st["C1_cw_04"]	= 0x2c
def_q_arg_st["C1_cw_05"]	= 0x2d
def_q_arg_st["C1_cw_06"]	= 0x2e

# Two-qubit operations using the 'T' registers
# Flux operations currently require a codeword from 128 to 255
def_q_arg_tt["fl_cw_01"]	= 0x81
def_q_arg_tt["fl_cw_00"]	= 0x80
def_q_arg_tt["fl_cw_02"]	= 0x82
def_q_arg_tt["fl_cw_03"]	= 0x83
def_q_arg_tt["fl_cw_04"]	= 0x84
def_q_arg_tt["fl_cw_05"]	= 0x85
def_q_arg_tt["fl_cw_06"]	= 0x86
def_q_arg_tt["fl_cw_07"]	= 0x87
\end{lstlisting}

%% file: ms.bbl
\begin{thebibliography}{1}
\providecommand{\url}[1]{#1}
\csname url@samestyle\endcsname
\providecommand{\newblock}{\relax}
\providecommand{\bibinfo}[2]{#2}
\providecommand{\BIBentrySTDinterwordspacing}{\spaceskip=0pt\relax}
\providecommand{\BIBentryALTinterwordstretchfactor}{4}
\providecommand{\BIBentryALTinterwordspacing}{\spaceskip=\fontdimen2\font plus
\BIBentryALTinterwordstretchfactor\fontdimen3\font minus
  \fontdimen4\font\relax}
\providecommand{\BIBforeignlanguage}[2]{{%
\expandafter\ifx\csname l@#1\endcsname\relax
\typeout{** WARNING: IEEEtran.bst: No hyphenation pattern has been}%
\typeout{** loaded for the language `#1'. Using the pattern for}%
\typeout{** the default language instead.}%
\else
\language=\csname l@#1\endcsname
\fi
#2}}
\providecommand{\BIBdecl}{\relax}
\BIBdecl

\bibitem{fu2018eqasm}
X.~Fu, L.~Riesebos, M.~A. Rol, J.~van Straten, J.~van Someren, N.~Khammassi,
  I.~Ashraf, R.~F.~L. Vermeulen, V.~Newsum, K.~K.~L. Loh, J.~C. de~Sterke,
  W.~J. Vlothuizen, R.~N. Schouten, C.~G. Almudever, L.~DiCarlo, and
  K.~Bertels, ``{eQASM: An Executable Quantum Instruction Set Architecture},''
  in \emph{Proceedings of 25th IEEE International Symposium on High-Performance
  Computer Architecture (HPCA’19)}.\hskip 1em plus 0.5em minus 0.4em\relax
  IEEE, 2019, pp. 224--237.

\bibitem{fu2017experimental}
X.~Fu, M.~A. Rol, C.~C. Bultink, J.~van Someren, N.~Khammassi, I.~Ashraf,
  R.~F.~L. Vermeulen, J.~C. de~Sterke, W.~J. Vlothuizen, R.~N. Schouten, C.~G.
  Almudever, L.~DiCarlo, and K.~Bertels, ``An experimental microarchitecture
  for a superconducting quantum processor,'' in \emph{Proceedings of the 50th
  Annual IEEE/ACM International Symposium on Microarchitecture
  (MICRO-50)}.\hskip 1em plus 0.5em minus 0.4em\relax ACM, 2017, pp. 813--825.

\end{thebibliography}
